\documentclass[final,5p,times,twocolumn]{elsarticle}
\usepackage{amssymb}
\usepackage{amsmath}
\usepackage{algorithmic}
\usepackage{tikz}
\usetikzlibrary{shapes.geometric, arrows,calc}

\tikzstyle{1d} = [rectangle, rounded corners, minimum width=2.5cm, minimum height=0.7cm, text centered, draw=black, fill=green!40, text width=6.5cm]
\tikzstyle{2d} = [rectangle, rounded corners, minimum width=2.5cm, minimum height=0.7cm, text centered, draw=black, fill=blue!20, text width=6.5cm]
\tikzstyle{2dsep} = [rectangle, rounded corners, minimum width=2.5cm, minimum height=0.7cm, text centered, draw=black, fill=yellow!50, text width=6.5cm]
\tikzstyle{collective} = [rectangle, rounded corners, minimum width=2.5cm, minimum height=0.7cm, text centered, draw=black, fill=red!50, text width=6.5cm]
\tikzstyle{decision} = [diamond,aspect=2,minimum width=1cm, minimum height=0.7cm, text centered, draw=black, fill=red!50]
\tikzstyle{arrow} = [thick,->,>=stealth]

\newcommand{\I}{{\mathrm i}}
\newcommand{\NB}{\mathit{NB}}
\newcommand{\MB}{\mathit{MB}}

\begin{document}
\begin{frontmatter}
\title{Scalable Nuclear Density Functional Theory with Sky3D}
\author[a]{Md Afibuzzaman}
\ead{afibuzza@msu.edu}
\author[b]{Bastian Schuetrumpf}
\ead{schutrum@nscl.msu.edu}
\author[a]{Hasan Metin Aktulga\corref{author}}
\ead{hma@cse.msu.edu}

\cortext[author] {Corresponding author}

\address[a]{Computer Science and Engineering, Michigan State University
  East Lansing, Michigan 48824, USA}
\address[b]{FRIB Laboratory Michigan State University East Lansing, Michigan 48824, USA}

\begin{abstract}
In nuclear astrophysics, quantum simulations of large inhomogeneous dense systems as they appear in the crusts of neutron stars present big challenges. The number of particles in a simulation with periodic boundary conditions is strongly limited due to the immense computational cost of the quantum methods. In this paper, we describe techniques for an efficient and scalable parallel implementation of Sky3D, a nuclear density functional theory solver that operates on an equidistant grid. Presented techniques allow Sky3D to achieve good scaling and high performance on a large number of cores, as demonstrated through detailed performance analysis on a Cray XC40 supercomputer. 
\end{abstract}

\end{frontmatter}

\section{Introduction}
Compact objects in space such as neutron stars are great test laboratories for nuclear physics as they contain all kinds of exotic nuclear matter which are not accessible in experiments on earth \cite{Bethe,Suzuki}. Among interesting kinds of astromaterial is the so called nuclear "pasta" phase \cite{Ravenhall,Hashimoto} which consists of neutrons and protons embedded in an electron gas. The naming arises from the shapes, e.g. rods and slabs, which resemble the shapes of the Italian pasta (spaghetti and lasagna). 

Nuclear pasta is expected in the inner crust of neutron stars in a layer of about 100\,m at a radius of about 10\,km, at sub-nuclear densities. Since the typical length scale of a neutron or a proton is on the order of 1\,fm, it is impossible to simulate the entire system. The usual strategy is to simulate a small part of the system in a finite box with periodic boundary conditions. While it is feasible to perform large simulations with semi-classical methods such as Molecular Dynamics (MD) \cite{Schneider13,Schneider14,Horowitz2015a,Watanabe2002,Watanabe:2003,Watanabe2005,Watanabe09} involving 50,000 or even more particles or the Thomas-Fermi approximation \cite{Williams1985,Oyamatsu1993,Oka13a,Pais15}, quantum mechanical (QM) methods which can yield high fidelity results have been limited to about 1000 nucleons due to their immense computational costs \cite{Mag02,Goe07a,NewtonStone,Sonoda2008,Schuetrumpf2014,Schuetrumpf2015a,Sagert2016,Fattoyev2017}.

The side effects of using small boxes in QM methods are twofold: First, the finite size of the box causes finite-volume effects, which have an observable influence on the results of a calculation. Those effects have been studied and can be suppressed by introducing the twist-averaged boundary conditions \cite{Schuetrumpf2015a}. More importantly though, finite boxes limit the possible resulting shapes of the nuclear pasta because the unit cell of certain shapes might be larger than the maximum box size. For instance, in MD simulations \cite{Horowitz2015}, slabs with certain defects have been discovered. Those have not been observed in QM simulations because they only manifest themselves in large boxes. To observe such defects, we estimate that it is necessary to increase the number of simulated particles (and the corresponding simulation box volume) by about an order of magnitude.

In this paper, we focus on the microscopic nuclear density functional theory (DFT) approach to study nuclear pasta formations. The nuclear DFT approach is a particularly good choice for nuclear pasta. The most attractive property is the reliability of its answers over the whole nuclear chart \cite{Erler2012,Bender03}, and yet it is computationally feasible for applications involving the heaviest nuclei and even nuclear pasta matter, because the interaction is expressed through one-body densities and the explicit n-body interactions do not have to be evaluated. 

In contrast to finite nuclei that are usually calculated employing a harmonic oscillator finite-range basis \cite{Stoitsov2013a,Schunck2012} using mostly complete diagonalization of the basis functions to solve the self-consistent equations, nuclear pasta matter calculations have to be performed in a suitable basis with an infinite range. We use the DFT code Sky3D\,\cite{Mar15a}, which represents all functions on an equidistant grid and employs the damped gradient iteration steps, where fast Fourier transforms (FFTs) are used for derivatives, to reach a self-consistent solution. This code is relatively fast compared to its alternatives and incorporates all features necessary to perform DFT calculations with modern functionals, such as the Skyrme functionals (as used here). Since Sky3D can be used to study static and time-dependent systems in 3d without any symmetry restrictions, it can be applied to a wide range of problems. In the static domain it has been used to describe a highly excited torus configuration of $^{40}$Ca \cite{ichikawa2012} and also finite nuclei in a strong magnetic field as present in neutron stars \cite{Stein}. In the time-dependent context, it was used for calculations on nuclear giant resonances \cite{Reinhard2007,Schuetrumpf2016}, and on the spin excitation in nuclear reactions \cite{Maruhn2006}. The Wigner function, a 6 dimensional distribution function, and numerical conservation properties in the time-dependent domain have also been studied using Sky3D \cite{Loebl2011, Guo2008}.

For the case of time-dependent problems, the Sky3D code has already been parallelized using MPI. The time-dependent iterations are simpler to parallelize, because the treatment of the single particles are independent of each other and can be distributed among the nodes. Only the mean field has to be communicated among the computational nodes. On the other hand, accurate computation of nuclear ground states, which we are interested in, requires a careful problem decomposition strategy and organization of the communication and computation operations as discussed below. However, only a shared memory parallel version of Sky3D (using OpenMP) exists to this date. In this paper, we present algorithms and techniques to achieve scalable distributed memory parallelism in Sky3D using MPI.
\section{Background}

\subsection{Nuclear Density Functional Theory with Skyrme Interaction}
Unlike in classical calculations where a point particle is defined by its position and its momentum, quantum particles are represented as complex wave functions. The square modulus of the wave function in real space is interpreted as a probability amplitude to find a particle at a certain point. Wave functions in the real space and the momentum space are related via the Fourier transform. In the Hartree-Fock approximation used in  nuclear DFT calculations, the nuclear N-body wave function is restricted to a single Slater determinant consisting of N orthonormalized one-body wave functions $\psi_\alpha$, $\alpha=1..N$. Each of these one-body wave functions have to fulfill the one-body Schr\"odinger's Equation
\begin{equation}
\hat{h}_q\psi_\alpha=\epsilon_\alpha\psi_\alpha, \label{eq:Schroedinger}
\end{equation}
when convergence is reached, \emph{i.e.}, when 
\begin{eqnarray}
    \overline{\Delta\varepsilon}
    &=&
    \sqrt{\frac{\sum_\alpha\langle\psi_\alpha|\hat{h}^2|\psi_\alpha\rangle
    -
    \langle\psi_\alpha|\hat{h}|\psi_\alpha\rangle^2}{\sum_\alpha 1}}\label{eq:convergence}
  \end{eqnarray}
is small. 

In nuclear DFT, the interaction between nucleons (\emph{i.e.}, neutrons and protons) is expressed through a mean field. In this work, we utilize the Skyrme mean field \cite{Bender03}:
\begin{align}
\mathcal{E}_{\rm Sk}&=\sum_{q=n,p}\left(C_q^\rho(\rho_0)\rho_q^2+C_q^{\Delta\rho}\rho_q\Delta\rho_q\right.\nonumber\\
&+\left.C_q^{\tau}\rho_q\tau_q+
C_q^{\nabla\vec{J}}\rho_q\nabla\vec{J}_q\right),\label{eq:Skyrme}
\end{align}
where the parameters $C_q^i$ have to be fitted to experimental observables. The mean field is determined by nucleon densities and their derivatives:
\begin{subequations}
\begin{align}
   \rho_q(\vec r)&=\displaystyle
    \sum_{\alpha\in q}\sum_{s}  
    v_{\alpha}^2|\psi_{\alpha}(\vec{r},s)|^2 \label{eq:rho}\\
   \vec{J}_q(\vec r) &=\displaystyle
    -\I\sum_{\alpha\in q}\sum_{ss'} v_{\alpha}^2
    \psi_{\alpha}^*(\vec{r},s)
    \nabla\! \times\! \vec{\sigma}_{ss'} 
    \psi^{\mbox{}}_{\alpha}(\vec{r},s')\label{eq:curr}\\
    \tau_q(\vec r)&=\displaystyle
    \sum_{\alpha\in q}\sum_{s}  
    v_{\alpha}^2|\nabla\!\psi_{\alpha}(\vec{r},s)|^2\quad ,
 \label{eq:tau}
\end{align}
\end{subequations}
where $\rho_q$ is the number density, $\vec{J}_q$ is the spin-orbit density and $\tau_q(\vec r)$ is the kinetic density for $q\in$ (protons,neutrons), which are calculated from the wave functions. We assume a time-reversal symmetric state in the equations. The interaction is explicitly isospin dependent. The parameters $v_\alpha^2$ are either 0 for non-occupied states or 1 for occupied states for calculations without the pairing force. With those occupation probabilities, the calculation can also be performed using more wave functions than the number of particles. The sum $\sum_\alpha v^2_\alpha$ determines the particle number. A detailed description of the Skyrme energy density functional can be found in references \cite{Kortelainen2012,Kluepfel2009}.

In DFT, the ground states associated with a many-body system is found in a self-consistent way, \emph{i.e.}, iteratively. The self-consistent solution can be approached through the direct diagonalization method or, in this case we use the damped gradient iteration method which is described below. While stable finite nuclei as present on earth typically do not contain more than a total of 300 nucleons, nuclear pasta matter in neutron stars is quasi-infinite on the scales of quantum simulations. Therefore it is desirable to simulate as large volumes as possible to explore varieties of nuclear pasta matter. Furthermore, in contrast to finite nuclei which are approximately spherical, pasta matter covers a large range of shapes and deformations and thus many more iterations are needed to reach convergence. Since larger volumes and consequently more nucleons require very intensive calculations, a high performance implementation of nuclear DFT codes is desirable.

DFT is also widely used for electronic structure calculations in computational chemistry. While DFT approaches used in computational chemistry can efficiently diagonalize matrices associated with a large number of basis sets, we need to rely on different iteration techniques in nuclear DFT. The most important reason for this is that in computational chemistry, electrons are present in a strong external potential. Therefore, iterations can converge relatively quickly in this case. However, in nuclear DFT, the problem must be solved in a purely self-consistent manner because nuclei are self-bound. As a result, the mean field can change drastically from one iteration to the next, since no fixed outer potential is present. Especially for nuclear pasta spanning a wide range of shapes, a few thousand iterations are necessary for the solver to converge. Therefore, nuclear DFT iterations have to be performed relatively quickly, making it infeasible to employ the electronic DFT methods which are expensive for a single iteration.

\subsection{Sky3D Software}
Sky3D is a nuclear DFT solver, which has frequently been used for finite nuclei, as well as for nuclear pasta (for both static and time-dependent) simulations. The time-dependent version of Sky3D is relatively simpler to parallelize compared to the static version, because properties like orthornomality of the wave functions are implicitly conserved due to the fact that the time-evolution operator is unitary. Therefore the calculation of a single nucleon is independent of the others. The only interaction between nucleons takes place through the mean field. Thus only the nuclear densities using which the mean field is constructed has to be communicated. In the static case, however, orthonormality has to be ensured and the Hamiltonian matrix must be diagonalized to obtain the eigenvalues and eigenvectors of the system at each iteration. In this paper, we describe parallelization of the more challenging static version (which previously was only shared memory parallel).

Sky3D operates on a three dimensional equidistant grid in coordinate space. Since nuclear DFT is a self-consistent method requiring an iterative solver, the calculation has to be initialized with an initial configuration. In Sky3D, the wave functions are initialized with a trial state, using either the harmonic oscillator wave functions for finite nuclei or plane waves for periodic systems. The initial densities and the mean field are calculated from those trial wave functions.

After initialization, iterations are performed using the damped gradient iteration scheme \cite{Blum1992}
\begin{equation}
  \psi_\alpha^{(n+1)}
  =
  \mathcal{O}\left\{
    \psi_\alpha^{(n)}
    -
    \frac{\delta}{\hat{T} + E_0} 
    \left( \hat{h}^{(n)} - 
      \langle\psi_\alpha^{(n)}|\hat{h}^{(n)}|\psi_\alpha^{(n)}\rangle
    \right)\psi_\alpha^{(n)}\right\}\quad,
  \label{eq:dampstep}
\end{equation}
where $\mathcal{O}$ denotes the orthonormalization of the wave functions, $\hat{T}$ denotes the kinetic energy operator, $\psi_\alpha^{(n)}$ and $\hat{h}^{(n)}$ denote the single-particle wave function and the Hamiltonian at step $n$, respectively, and $\delta$ and $E_0$ are constants that need to be tuned for fast convergence. The Hamiltonian consists mainly of the kinetic energy, the mean field contribution (Eq.\ref{eq:Skyrme}) and the Coulomb contribution. We use FFTs to compute the derivatives of the wave functions. The Coulomb problem is solved in the momentum space, also employing FFTs.

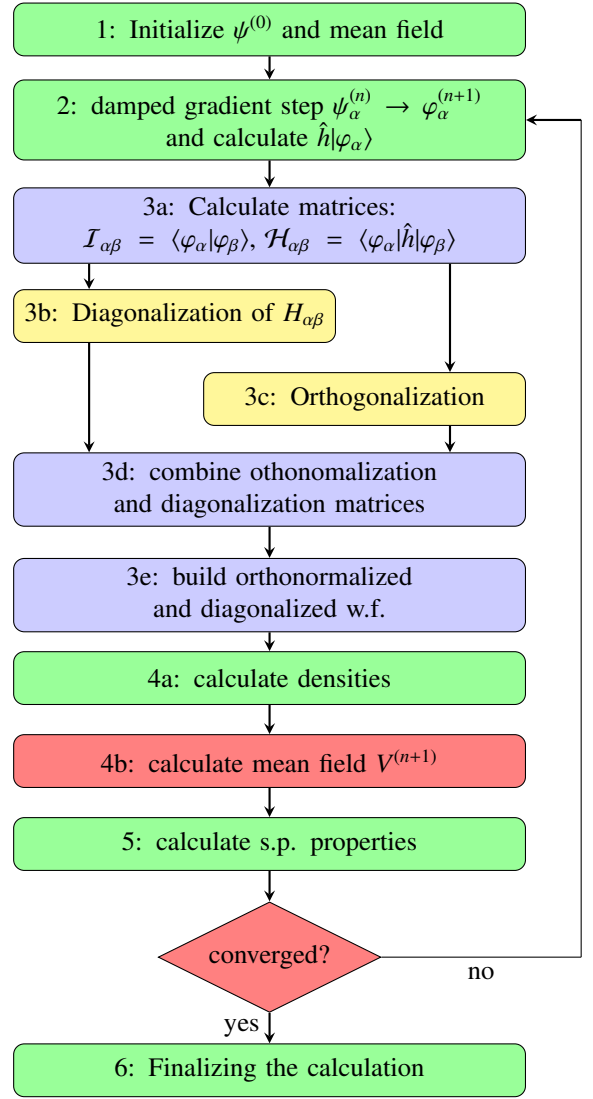
\begin{figure}
\centering
\begin{tikzpicture}[node distance=1.1cm]
\node (init) [1d] {1: Initialize $\psi^{(0)}$ and mean field};
\node (grstep) [1d, below of=init,yshift=-0.1cm] {2: damped gradient step $\psi_\alpha^{(n)} \rightarrow \varphi_\alpha^{(n+1)}$\\
and calculate $\hat{h}|\varphi_\alpha\rangle$};
\node(calcmatrix)[2d, below of=grstep,yshift=-0.3cm]{3a: Calculate matrices:\\ $\mathcal{I}_{\alpha\beta}  = \langle\varphi_\alpha|\varphi_\beta\rangle$,
$\mathcal{H}_{\alpha\beta}=\langle\varphi_\alpha|\hat{h}|\varphi_\beta\rangle$};
\node(diagstep)[2dsep, below of=calcmatrix,text width=4cm,xshift=-1.25cm,yshift=-0.1cm]{3b: Diagonalization of $H_{\alpha \beta}$};
\node(GS3)[2dsep, below of=diagstep,text width=4cm,xshift=2.5cm]{3c: Orthogonalization};
\node(combine)[2d, below of=GS3,xshift=-1.25cm,yshift=-0.1cm]{3d: combine othonomalization and diagonalization matrices};
\node(recombine)[2d, below of=combine,yshift=-0.3cm]{3e: build orthonormalized\\ and diagonalized w.f.};
\node(dens)[1d, below of=recombine]{4a: calculate densities};
\node(skyrme)[collective, below of=dens]{4b: calculate mean field $V^{(n+1)}$};
\node(sp_prop)[1d, below of=skyrme]{5: calculate s.p. properties};
\node(converged)[decision, below of=sp_prop,yshift=-0.4cm]{converged?};
\node(stop)[1d, below of=converged,yshift=-0.4cm]{6: Finalizing the calculation};
\node(dummy1)[right of=converged,xshift=3cm]{};
\node(dummy2)[right of=grstep,xshift=3cm]{};
\draw [arrow] (init)   -- (grstep);
\draw [arrow] (grstep) -- (calcmatrix);
\draw [arrow] ($(calcmatrix.south west)+(1,0)$) -- ($(diagstep.north west)+(1,0)$);
\draw [arrow] ($(calcmatrix.south east)-(1,0)$) -- ($(GS3.north east)-(1,0)$);
\draw [arrow] ($(diagstep.south west)+(1,0)$) -- ($(combine.north west)+(1,0)$);
\draw [arrow] ($(GS3.south east)-(1,0)$) -- ($(combine.north east)-(1,0)$);
\draw [arrow] (combine) -- (recombine);
\draw [arrow] (recombine) -- (dens);
\draw [arrow] (dens) -- (skyrme);
\draw [arrow] (skyrme) -- (sp_prop);
\draw [arrow] (sp_prop) -- (converged);
\draw [arrow] (converged) -- node[anchor=east] {yes} (stop);
\draw (converged) -- node[anchor=north] {no} (dummy1.center);
\draw (dummy1.center) -- (dummy2.center);
\draw [arrow] (dummy2.center) -- (grstep);
\end{tikzpicture}
\caption{Flowchart of the parallelized Sky3D code. Parts in the 1d distribution are marked in green, parts in full 2d distribution are marked in blue, parts in divided 2d distribution are marked in yellow and collective non-parallelized parts are marked in red. $\psi^{(n)}_\alpha$ denotes the orthonormal and diagonal wave function at step n with index $\alpha$, $|\varphi_\alpha\rangle=\varphi_\alpha^{(n+1)}$ denotes the non-diagonal and non-orthonormal wave function at step n+1. $\hat{h}$ is the one-body Hamiltonian.
\label{fig:Sky3d_para}}
\end{figure}

The basic flow chart of the static Sky3D code is shown in Fig~\ref{fig:Sky3d_para}. Since the damped gradient iterations of Eq.~\ref{eq:dampstep} does not conserve the diagonality of the wave functions with respect to the Hamiltonian, i.e. $\langle\psi_\alpha|\hat{h}|\psi_\beta\rangle=\delta_{\alpha\,\beta}$, they have to be diagonalized  after each step to obtain the eigenfunctions. Subsequently, single-particle properties, \emph{e.g.} single-particle energies, are determined. If the convergence criterion (Eq.(\ref{eq:convergence})) is fulfilled at the end of the current iteration, properties of the states and the wave functions are written into a file and the calculation is terminated.

\section{Distributed Memory Parallelization with MPI}
There are basically two approaches for distributed memory parallelization of the Sky3D code. The first approach would employ a spatial decomposition where the three dimensional space is partitioned and corresponding grid points are distributed over different MPI ranks. However, computations like the calculation of the density gradients $\nabla \rho_q(\vec{r})$ are global operations that require 3D FFTs, which are known to have poor scaling due to their dependence on all-to-all interprocess communications \cite{sunderland2012analysis}. Hence, this approach would not scale well. The second approach would be to distribute the single particle wave functions among processes. While communications are unavoidable, by carefully partitioning the data at each step, it is possible to significantly reduce the communication overheads in this scheme. In what follows, we present our parallel implementation of Sky3D using this second approach. 

The iterative steps following the initialization phase constitute the computationally expensive part of Sky3D. Hence, our discussion will focus on the implementation of steps 2 through 5 of Fig.~\ref{fig:Sky3d_para}. Computations in these steps can be classified into two groups, i) those that work with matrices (and require 2D distributions to obtain good scaling), and ii) those that work on the wave functions themselves (and utilize 1D distributions as it is more convenient in this case to have wave functions to be fully present on a node). Our parallel implementation progresses by switching between these 2D partitioned steps (marked in violet and yellow in Fig.\,\ref{fig:Sky3d_para}) and 1D partitioned steps (marked in green) in each iteration. Steps marked in red are not parallelized.

As discussed in more detail below, an important aspect of our implementation is that we make use of optimized scientific computing libraries such as ScaLAPACK\,\cite{blackford1997scalapack} and FFTW\,\cite{frigo1998fftw} wherever possible. Since ScaLAPACK and FFTW are widely used and well optimized across HPC systems, this approach allows us to achieve high performance on a wide variety of architectures without the added burden of fine-tuning Sky3D. This may even extend to future architectures with decreased memory space per core and possibly multiple levels in the memory hierarchy. As implementations of ScaLAPACK and FFTW libraries evolve for such systems, we anticipate that it will be relatively easy to adapt our implementation to such changes in HPC systems.

\subsection{The 1D and 2D Partitionings}

The decisions regarding 1D and 2D decompositions are made around the wave functions which represent the main data structure in Sky3D and are involved in all the key computational steps. We represent the wave functions using a two dimensional array {\tt psi(V,A)}, where $V=n_x\times n_y\times n_z\times 2$ includes the spatial degrees of freedom and the spin degree of freedom with $n_x$, $n_y$ and $n_z$ being the grid sizes in $x$, $y$ and $z$ directions, respectively, and the factor 2 originating from the two components of the spinor. In the case of 1D distribution, full wave functions are distributed among processes in a block cyclic way. The block size $N_\psi$ determines how many consecutive wave functions are given to each process in each round. In round one, the first $N_\psi$ wave functions are given to the first process $P_0$, then the second process $P_1$ gets the second batch and so on. When all processes are assigned a batch of wave functions in a round, the distribution resumes with $P_0$ in the subsequent round until all wave functions are exhausted. 

\begin{figure}[!t]
\centering
\includegraphics[width=\columnwidth]{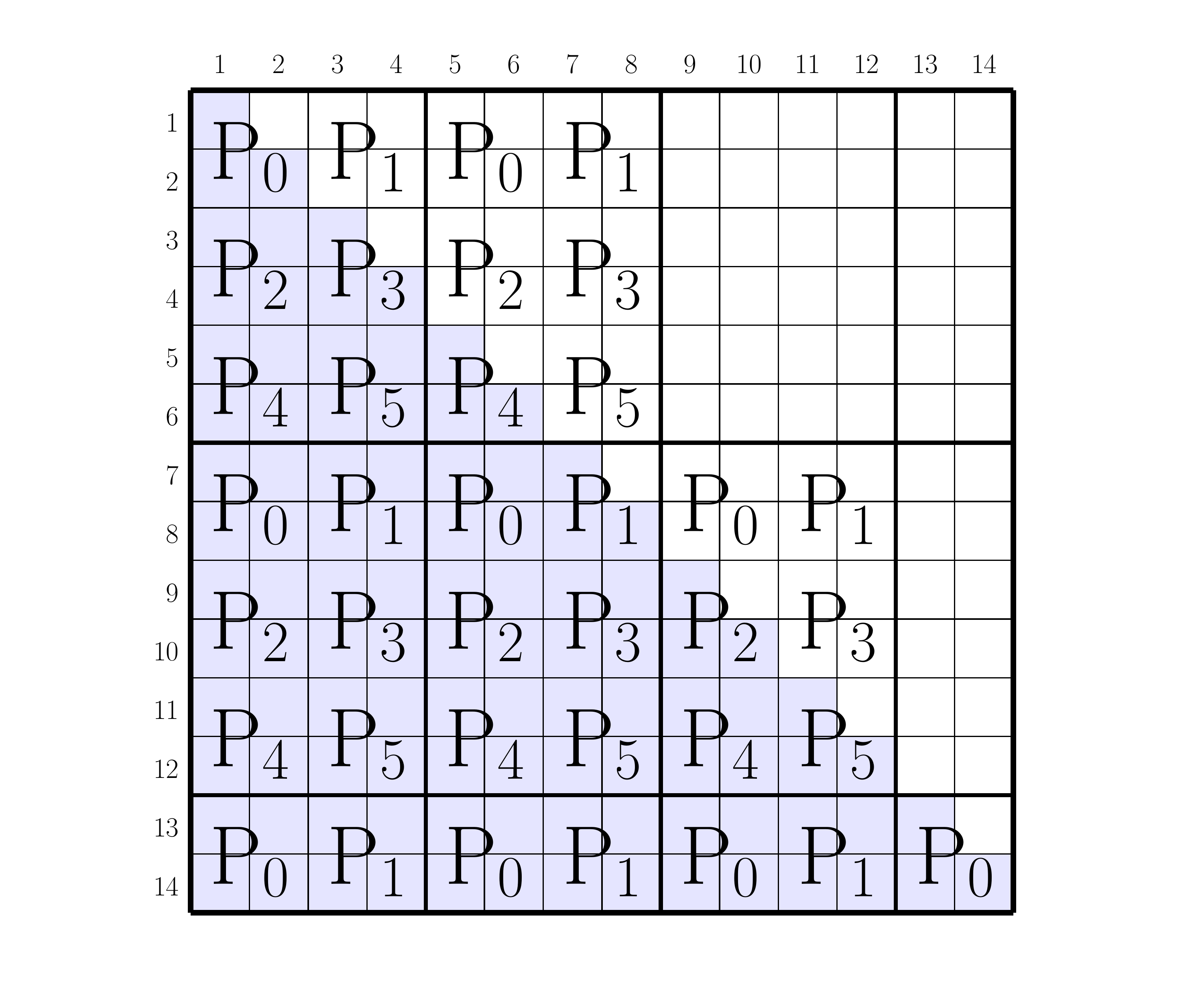}
\caption{2D block cyclic partitioning example with 14 wave functions using a 3$\times$2 processor topology. Row and column block sizes are set as $\NB$=$\MB$=2. The blue shaded area marks the lower triangular part. }
\label{fig:2dcycl}
\end{figure}

In the 2D partitioning case, single particle wave functions as well as the matrices constructed using them (see Sect.\,\ref{sec:matcons}) are divided among processes using a 2D block cyclic distribution. In Fig.~\ref{fig:2dcycl}, we provide a visual example of a square matrix distributed in a 2D block cyclic fashion where processes are organized into a 3$\times$2 grid topology, and the row block size $\NB$ and the column block size $\MB$ have been set equal to 2. The small rectangular boxes in the matrix show the arrangement of processes in the 3$\times$2 grid topology -- the number of rows are in general not equal to the number of columns in the process grid. For symmetric or Hermitian matrices only the (blue marked) lower triangular part is needed as it defines the matrix fully. In this particular case, $P_0$ is assigned the matrix elements in rows 1, 2, 7, 8, 13 and 14 in the first column, as well as those in rows 2, 7, 8, 13 and 14 in the second column; $P_1$ is assigned the matrix elements in rows 7, 8, 13 and 14 in columns 3 and 4, and so on. Single particle wave functions, which are stored as rectangular (non-symmetric) matrices with significantly more number of rows than the number of columnd (due to the large grid sizes needed), are also distributed using the same strategy. The 2D block cyclic distribution provides very good load balance in computations associated with the single particle wave functions and the matrices formed using them, as all processes are assigned approximately the same number of elements.

\subsection{Parallelization across Neutron and Proton Groups}
An important observation for parallelization of Sky3D is that neutrons and protons interact only through the mean field. Therefore, the only communication needed between these two species takes place within step 4b. To achieve better scalability, we leverage this fact and separate computations associated with neutrons and protons to different processor groups, while trying to preserve the load balance between them as explained in detail below.

Suppose $A=N+Z$ is the total number of wave functions, such that N is the number of neutron wave functions and Z is the number of proton wave functions. To distribute the array {\tt psi}, we split the available processors into two groups; one group of size $P_N$ for neutrons, and another group of size $P_P$ for protons. We note that in practical simulations, the number of neutrons are often larger than the number of protons (for example, simulations of neutron star matter are naturally neutron rich). The partitioning of processors into neutron and proton groups must account for this situation to ensure good load balance between the two groups. As will be discussed in Section\,\ref{sec:2D}, Sky3D execution time is mainly determined by the time spent on 2D partitioned steps which is primarily dominated by the construction of the overlap and Hamiltonian matrices step, and to a lesser degree by eigenvalue computations associated with these matrices. Since the computational cost of the matrix construction step is proportional to the square of the number of particles in a group (see Section\,\ref{sec:matcons}), we choose to split processors into two groups by quadratically weighing the number of particles in each species. More precisely, if the total processor count is given by $C$, then $P_N = \frac{N^2}{N^2+Z^2}C$ and $P_P = \frac{Z^2}{N^2+Z^2}C$ according to this scheme. It is well-known that 2D partitioning is optimal for scalable matrix-matrix multiplications\,\cite{van1997summa}. Therefore, once the number of processors within neutron and proton groups is set, we determine the number of rows and columns for the 2D process topologies of each of these groups through MPI's {\tt MPI\_DIMS\_CREATE} function. This function ensures that the number of rows and columns is as close as possible to the square root of $P_N$ and $P_P$ for neutron and proton groups, respectively, thus yielding a good 2D process layout.

As will be demonstrated through numerical experiments, the scheme described above typically delivers computations with good load balances, but it has a caveat. Under certain circumstances, this division might yield a 2D process grid with a tall \&  skinny layout, which essentially is more similar to a 1D partitioning and have led to significant performance degradations for our 2D partitioned computations. For instance, in a system with 5000 neutrons and 1000 protons, when we use 256 processors, according to our scheme $P_N$ will be 246, and $P_P$ will be 10. For $P_N = 246$, the corresponding process grid is $(41*6)$ which is much closer to a 1D layout than a 2D layout. To prevent such circumstances, we require the number of processors within each group to be a multiple of certain powers of 2 (\emph{i.e.}, 2, 4, 8, 16 or 32 depending on the total core count). For the above example, by requiring the number of cores within each group to be a multiple of 16, we determine $P_N$ to be 240 and $P_P$ to be 16. This results in a process grid of size $16\times15$ for neutrons which is almost a square shaped grid, and a perfect square grid of size $4\times4$ for protons. 

\subsection{Calculations with 2D Distributions}
\label{sec:2D}
As a result of the split, neutron and proton processor groups asynchronously advance through steps that require 2D partitionings, \emph{i.e.}, steps 3a to 3e. Main computational tasks here are the construction of the overlap and Hamiltonian matrices (using 1D distributed wave functions) and eigendecompositions of these matrices. These tasks are essentially accomplished by calling suitable ScaLAPACK routines. The choice of a 2D block cyclic distribution maximizes the load balancing with ScaLAPACK for the steps 3a through 3e.
\subsubsection{Matrix Construction (step 3a)}
\label{sec:matcons}
Construction of the overlap matrix $\mathcal{I}$ and the Hamiltonian matrix $\mathcal{H}$ 
\begin{eqnarray}
\mathcal{I}_{\alpha\beta}&=&\langle\varphi_\alpha|\varphi_\beta\rangle, \label{eq:normmatrix} \mbox{  and}\\
\mathcal{H}_{\alpha\beta}&=&\langle\varphi_\alpha|\hat{h}|\varphi_\beta\rangle, \label{eq:hammatrix}
\end{eqnarray}
where $|\varphi_\alpha\rangle$ marks the non-orthonormalized and non-diagonal wave functions, constitutes the most expensive part of the iterations in Sky3D, because the cost of these operations scales quadratically with the number of particles. More precisely, calculating these matrices costs $\mathcal{O}(I^2V)$, where $I\in\{N,Z\}$ is the number of wave functions. Since these two operations are simply inner products, we use ScaLAPACK's complex matrix matrix multiplication routine \texttt{PZGEMM} for constructing these two matrices. 

One sublety here is that prior to the start of steps with 2D partitionings, wave functions are distributed in a 1D scheme. To achieve good performance and scaling with \texttt{PZGEMM}, we first switch both wave functions $|\varphi_\alpha\rangle$ ({\tt psi}) and $\hat{h}|\varphi_\alpha\rangle$ ({\tt hampsi}) into a 2D cyclic layout which uses the same process grid created through the {\tt MPI\_DIMS\_CREATE} function. The {\tt PZGEMM} call then operates on these two matrices and the complex conjugate of ({\tt psi}). The resulting matrices $\mathcal{I}$ and $\mathcal{H}$ are distributed over the same 2D process grid as well.

Since $\mathcal{I}$ and $\mathcal{H}$ are square matrices, we set the row and column block sizes, $\mathit{NB}$ and $\mathit{MB}$, respectively, to be equal (\emph{i.e.,} $\mathit{NB}=\mathit{MB}$). Normally, in a 2D matrix computation, one would expect a trade-off in choosing the exact value for $\mathit{NB}$ and $\mathit{MB}$, as small blocks lead to a favorable load balance, but large blocks reduce communication overheads. However, for typical problems that we are interested in, \emph{i.e.}, more than 1000 particles using at least a few hundred cores, our experiments with different $\mathit{NB}$ and $\mathit{MB}$ values such as 2, 4, 8, 32, 64 have shown negligible performance differences. Therefore, we have empirically determined the choice for $\mathit{NB}=\mathit{MB}$ to be 32 for large computations.

\subsubsection{Diagonalization and Orthonormalization (steps 3b \& c)}

After the matrix $\mathcal{H}$ is constructed according to Eq.~(\ref{eq:hammatrix}), its eigenvalue decomposition is computed to find the eigenstates of the Hamiltonian. Since $\mathcal{H}$ is a hermitian matrix, we use the complex Hermitian eigenvalue decomposition routine \texttt{PZHEEVR} in ScaLAPACK, which first reduces the input matrix to tridiagonal form, and then computes the eigenspectrum using the Multiple Relatively Robust Representations (MRRR) algorithm\,\cite{dhillon2006design}. The local matrices produced by the 2D block cyclic distribution of the matrix construction step can readily be used as input to the \texttt{PZHEEVR} routine. After the eigenvectors of $\mathcal{H}$ are obtained, the diagonal set of wave functions $\psi_\alpha$ can be obtained through the following matrix-vector multiplication 
\begin{equation}
\psi_\alpha=\sum_\beta \mathcal{Z}^H_{\alpha\beta}\varphi_\beta
\end{equation}
for all $\varphi_\beta$ where $\mathcal{Z}$ is the matrix containing the eigenvectors of $\mathcal{H}$. 

Orthonormalization is commonly accomplished through the modified Gram-Schmidt (mGS) method, a numerically more stable version of the \emph{classical} Gram-Schmidt method. Unlike the original version of Sky3D, we did not opt for mGS for a number of reasons. 
First, mGS is an inherently sequential process where the orthonormalization of wave function $n+1$ can start only after the first $n$ vectors are orthonormalized. Second, the main computational kernel in this method is a dot product which is a Level-1 BLAS operation and has low arithmetic intensity. Finally, a parallel mGS with a block cyclic distribution of wave functions $|\varphi_\alpha\rangle$ as used by matrix construction and diagonalization steps would incur significant synchronization overheads, especially due to the small blocking factors needed to load balance the matrix construction step. 

An alternative approach to orthonormalize the wave functions is the L\"owdin method \cite{Loewdin1950}, which can be stated for our purposes as: 
\begin{subequations}
\begin{eqnarray}
\mathcal{C} &=& \mathcal{I}^{-1/2}\label{eq:Loewedin}\\
\psi_\alpha &=& \sum_{\beta}\mathcal{C}_{\beta\alpha} \varphi_\beta \label{eq:orthmatvec}.
\end{eqnarray}
\end{subequations}
The L\"owdin orthonormalization is well known in quantum chemistry and has the property that the orthornormalized wave functions are those that are closest to the non-orthonormalized wave functions in a least-squares sense. 
Note that since $\mathcal{I}$ is a Hermitian matrix, it can be factorized as $\mathcal{I} = X \Lambda X^T$, where columns of $X$ are its eigenvectors and $\Lambda$ is a diagonal matrix composed of $\mathcal{I}$'s eigenvalues. Consequently, $\mathcal{I}^{-1/2}$ in Eq.\ref{eq:Loewedin} can be computed simply by taking the inverses of the square roots of $\mathcal{I}$'s eigenvalues, \emph{i.e.}, 
$\mathcal{C}=\mathcal{I}^{-1/2}=X \Lambda^{-1/2} X^T$.

Applying the L\"owdin method in our problem is equivalent to computing an eigendecomposition of the overlap matrix $\mathcal{I}$, which can also be implemented by using the \texttt{PZHEEVR} routine in ScaLAPACK. Note that exactly the same distribution of wave functions and blocking factors as in the matrix construction step can be used for this step, too.

Detailed performance analyses reveal that the eigendecomposition routine \texttt{PZHEEVR} does not scale well for large $P$ with the usual number of wave functions in nuclear pasta calculations. However, the eigendecompositions of the $\mathcal{I}$ and $\mathcal{H}$ matrices (within both the neutron and proton groups) are independent of each other and their construction is also similar with respect to each other. Therefore, to gain additional performance, we perform steps 3b and 3c in parallel using half the number of MPI ranks available in a group, \emph{i.e.}, $P_N/2$ and $P_P/2$, respectively for neutrons and protons. 

\subsubsection{Post-processing (steps 3d \& e)}

The post-processing operations in the diagonalization and orthonormalization steps are matrix-vector multiplications acting on the current set of wave functions $\varphi_j$. As opposed to applying these operations one after the other, \emph{i.e.}, $\mathcal{C}^T(\mathcal{Z}^H\{\varphi\})$, we combine diagonalization and orthonormalization by performing $(\mathcal{C}^T\mathcal{Z}^H)\{\varphi\}$, where $\{\varphi\}=(\varphi_1,\varphi_2,..,\varphi_n)^T$ denotes a vector containing the single-particle wave functions. While both sequences of operations are arithmetically equivalent, the latter has a benefit in terms of the computational cost, as it reduces the number of multiply-adds from $2I^2V$ to $I^3 + I^2V$. This is almost half the cost of using the first sequence of operations, since we have $I<<V$ for both neutrons and protons, because the number of wave functions has to be significantly smaller than the size of the basis to prevent any bias due to the choice of the basis. We describe this optimization in the form of a pseudocode in Fig.~\ref{fig:step67}. By computing the overlap matrix $\mathcal{I}$ together with the Hamiltonian matrix $\mathcal{H}$, and performing their eigendecompositions, we can combine the update and orthonormalization of wave functions. 
Lines shown in red in Fig.~\ref{fig:step67} mark those affected by this optimization. Consequently, the matrix-matrix multiplication $C^T\mathcal{Z}^H$ can be performed prior to the matrix-vector multiplication involving the wave functions $\{\varphi(\vec{r}_\nu)\}$. As a result, the overall computational cost is significantly reduced, and efficient level-3 BLAS routines can be leveraged. 

The $C^T\mathcal{Z}^H$ operation is carried out in parallel using ScaLAPACK's \texttt{PZGEMM} routine (step 3d). Then the resulting matrix is multiplied with the vector of wave functions for all spatial and spin degrees of freedom in step 3e using another \texttt{PZGEMM} call.
\begin{figure}
\begin{algorithmic}[1]
\FOR{$q\in$ neutrons, protons}
\FOR{$i=1$ \TO $I$}
\FOR{$j=1$ \TO $I$}
\IF{$i\geq j$}
\STATE $\mathcal{H}_{ij}=\langle\varphi_i|\hat{h}|\varphi_j\rangle$
{\color{red}\STATE $\mathcal{I}_{ij}=\langle\varphi_i|\varphi_j\rangle$}
\ENDIF
\ENDFOR
\ENDFOR
\STATE Calculate $\mathcal{Z}$
{\color{red}\STATE Calculate $C$}
\FORALL{$\vec{r}_\nu$}
\STATE $\{\psi(\vec{r}_\nu)\}={\color{red}(C^T}\mathcal{Z}^H{\color{red})}\{\varphi(\vec{r}_\nu)\}$
\ENDFOR
\ENDFOR
\end{algorithmic}
\caption{Diagonalization and orthonormalization, which are steps 3a-b and 4 in the sequential version of Sky3D, are combined in the optimized parallel Sky3D code and are part of a single step (\emph{steps 3b-e}). In this pseudocode, $N_q$ labels the number of wave functions of the species q (proton or neutron) and $\{\varphi(\vec{r}_\nu)\}$ and $\{\psi(\vec{r}_\nu)\}$ are the vectors corresponding to the ``non-orthogonalized and non-diagonalized" and ``orthogonalized and diagonalized" wave functions, respectively. $\vec{r}_\nu$ labels the discretized position and spin degrees of freedom. Lines shown in red marks those affected by this optimization.}
\label{fig:step67}
\end{figure}

It should be noted that with this method, we introduce small errors during iterations, because the matrix $\mathcal{C}^T$ is calculated from non-orthogonalized wave functions. However, since we take small gradient steps, these errors disappear as we reach convergence and we ultimately arrive at the same results as an implementation which computes the matrix $\mathcal{C}^T$ from orthogonalized wave functions.

\subsection{Calculations with a 1D Distribution}

Steps 2, 4 and 5 use the 1D distribution, because for a given wave function $\psi_\alpha$, computations in these steps are independent of all other wave functions. Within the neutron and proton processor groups, we distribute wave functions using a 1D block cyclic distribution with block size $\NB_\psi$. Such a distribution further ensures good load balance and facilitates the communication between 1D and 2D distributions. The damped gradient step (as shown in the curled brackets in Eq.~\ref{eq:dampstep}) is performed in step 2. Here, the operator $\hat{T}$ shown in Eq.~\ref{eq:dampstep} is calculated using the FFTW library. Since the Hamiltonian has to be applied to the wave functions in this step, $\hat{h}|\psi_\alpha\rangle$ is saved in the array {\tt hampsi}, distributed in the same way as {\tt psi} and will be reused in step 3a. 
In step 4a, the partial densities as given in Eqs.~(\ref{eq:rho})-(\ref{eq:tau}) are calculated on each node for the local wave functions separately and subsequently summed up with the MPI routine \texttt{MPI\_ALLREDUCE}. We use FFTs to compute derivatives of wave functions as needed in Eqs. (\ref{eq:rho})-(\ref{eq:tau}).
The determination of the mean field in step 4b does not depend on the number of particles, and is generally not expensive. Consequently, this computation is performed redundantly by each MPI rank to avoid synchronizations. Also the check for convergence is performed on each MPI rank separately. Both are marked in red in Fig.~\ref{fig:Sky3d_para}.
Finally, in step 5, single-particle properties are calculated and partial results for single-particle properties are aggregated on all MPI ranks using an \texttt{MPI\_ALLREDUCE}.

\subsection{Switching between Different Data Distributions} 

As described above, our parallel Sky3D implementation uses 3 different distributions: The 1D distribution which is defined separately for neutrons and protons and used in green marked steps of Fig~\ref{fig:Sky3d_para}, the 2d distribution which is again defined separately for neutrons and protons and used in blue marked steps, and the 2d distribution for diagonalization and orthogonalization (used in steps marked in yellow) within subgroups of size $P_N/2$ and $P_P/2$, respectively for neutrons and protons. For calculation of overlap and Hamiltonian matrices, wave functions {\tt psi} and {\tt hampsi} need to be switched from the 1D distribution into the 2D distribution after step 2. After step 3a, matrices $\mathcal{I}$ and $\mathcal{H}$ must be transfered to the subgroups. After eigendecompositions in steps 3b and 3c are completed, the matrices $\mathcal{Z}$ and $\mathcal{C}$, which contain the eigenvectors, need to be redistributed back to the full 2D groups. After step 3e, only the updated array {\tt psi} has to be switched back to the 1D distribution from the 2D distributions. 

While these operations require complicated interprocess communications, they are easily carried out with the ScaLAPACK routine {\tt PZGEMR2D} which can transfer distributed matrices from one processor grid to another, even if the involved grids are totally different in their shapes and formations. As we will demonstrate in the performance evaluation section, the time required by {\tt PZGEMR2D} is insignificant, including in large scale calculations.

\subsection{Memory considerations}

Beyond performance optimizations, memory utilization is of great importance for large-scale nuclear pasta calculations. Data structures that require major memory space in Sky3D are the wave functions stored in matrices {\tt psi} and {\tt hampsi}. The latter matrix was not needed in the original Sky3D code as $\mathcal{H}$ was calculated on the fly, but this is not an option for a distributed memory implementation. As such, the total memory need increases by roughly a factor of 2. Furthermore, we store both arrays in both 1D and 2D distributions, which contributes another factor of 2. Besides the wavefunctions, another major memory requirement is storage of the matrices such as $\mathcal{H}$ and $\mathcal{I}$. These data structures are much smaller because the total matrix size grows only as $N^2$ for neutrons and $Z^2$ for protons. In our implementation, we store these matrices twice for the 2D distribution and twice for the 2D distributions within subgroups.

To give an example as to the actual memory utilization, largest calculations we conducted in this work are nuclear pasta calculations with a cubic lattice of 48 points and 6000 wave functions. In this case, the aggregate size of a single matrix to store wave functions in double precision complex format is $48^3\times 2\times 6000\times 16 \approx 21$\,GB, and all four arrays required in our parallelization would amount to about 84\,GBs of memory. The Hamiltonian and overlap matrices occupy a much smaller footprint, roughly 144 MBs per matrix. 
As this example shows, it is still feasible to carry out large scale nuclear pasta formations using the developed parallel Sky3D code. Even if we choose a bigger grid, \emph{e.g.}, of size $64^3$, with typical compute nodes in today's HPC systems having $\ge64$ GB of memory and the memory need per MPI rank decreasing linearly with the total number of MPI ranks in a calculation (there is little to no duplication of data structures across MPI ranks), such calculations would still be feasible using our implementation.

\section{Shared Memory Parallelization with OpenMP}

In addition to MPI parallelization, we also implemented a hybrid MPI/OpenMP parallel version of Sky3D. The rational behind a hybrid parallel implementation is that it would map more naturally to today's multi-core architectures, and it may also reduce the amount of inter-node communication using MPI. 

For the 1D distribution calculations, similar to the MPI implementation, we distribute the wave functions over threads by parallelizing loops using OpenMP. Since no communication is needed for step 2, we do not expect a gain in performance as a result of the shared memory implementation in this step. Step 4a, however, involves major communications, as the partial sums of the densities have to be reduced across all threads. The OpenMP implementation reduces the number of MPI ranks when the total core count $P$ is kept constant. This reduces the amount of inter-node communication. Similarly, step 5 involves the communication of single particle properties. Since these quantities are mainly scalars or small sized vectors though, inter-node communications are not as expensive.

The steps with a 2D distribution, \emph{i.e.}, steps 3a-3e, largely rely on ScaLAPACK routines. In this part, shared memory parallelization is introduced implicitly via the usage of multi-threaded ScaLAPACK routines. Consequently, note that we rely mostly on the ScaLAPACK implementation and its thread optimizations for the steps with 2D data distributions.

\section{Performance Evaluation}
\label{sec:performance}

\subsection{Experimental setup}

We conducted our computational experiments on Cori - Phase I, a Cray XC40 supercomputing platform at NERSC, which contains two 16-core Xeon E5-2698 v3 Haswell CPUs per node (see Table\,\ref{tab:machines}). Each of the 16 cores runs at 2.3\,GHz and is capable of executing one fused multiply-add (FMA) AVX (4$\times$64-bit SIMD) operation per cycle. Each core has a 64\,KB L1 cache (32\,KB instruction and 32\,KB data cache) and a 256\,KB L2 cache, both of which are private to each core. In addition, each CPU has a 40\,MB shared L3 cache. The Xeon E5-2698 v3 CPU supports hyperthreading which would essentially allow the use of 64 processes or threads per Cori-Phase I node. Our experiments with hyperthreading have led to performance degradation for both the MPI and MPI/OpenMP hybrid parallel implementations. As such, we have disabled hyperthreading in our performance tests.

{
\setlength{\tabcolsep}{5pt}

\begin{table}[!tb]{
\small
\centering
\renewcommand{\arraystretch}{1.5}
\begin{tabular}{rc}
\hline
{\bf Platform}		& {\bf Cray XC40} \\
{\bf Processor}		& {\bf Xeon E5-2698 v3} \\
\hline
{\bf Core}			& {\bf Haswell}	\\
\hline
Clock (GHz)			& 2.3			\\
Data Cache (KB)		& 64(32+32)+256		\\
Memory-Parallelism	& HW-prefetch	\\
\hline
Cores/Processor		& 16		\\
Last-level L3 Cache	& 40~MB		\\
SP TFlop/s			& 1.2		\\
DP TFlop/s			& 0.6		\\
STREAM BW$^3$		& 120\,GB/s	\\
Available Memory/node 	& 128\,GB 	\\
\hline
Interconnect		& Cray Aries (Dragonfly) \\
Global BW           & 5.625 TB/s \\
\hline
MPI Library			& MPICHv2 		\\
Compiler			& Intel/17.0.2.174	\\
\hline
\end{tabular}

\caption{\small 
Hardware specifications for a single socket on Cori, a Cray XC40 supercomputer at NERSC. Each node consists of two sockets. 
}
\label{tab:machines}
}
\vspace{-.15in}
\end{table}
}

For performance analysis, we choose a physically relevant setup. In practice, the grid spacing is about $\Delta x = \Delta y = \Delta z \sim 1 {\rm \,fm}$. This grid spacing gives results with desired accuracy \cite{Fattoyev2017}. For nuclear pasta matter, very high mean number densities ($0.02{\rm \,fm^{-3}}-0.14 {\rm \,fm^{-3}}$) have to be reached. We choose two different cubic boxes of $L=32\,{\rm fm}$ and $L=48\,{\rm fm}$ and a fixed number of nucleons $N+Z=6000$. This results in mean densities of $0.122{\rm \,fm^{-3}}$ and $0.036{\rm \,fm^{-3}}$, respectively.

To eliminate any load balancing effects, we begin with our performance and scalability tests using a symmetric system with 3000 neutrons and 3000 protons. 
As the systems for neutron star applications are neutron rich, we also test systems with 4000 neutron and 2000 protons and also 5000 neutrons and 1000 protons. 

\subsection{Scalability}

First, we consider a system with 3000 neutrons and 3000 protons with two different grid sizes, $L=32\,{\rm fm}$ and  $L=48\,{\rm fm}$. On Cori, each ``Haswell" node contains 32 cores on two sockets. For both grid sizes, we ran simulations on 1, 2, 4, 8, 16, 32 and 48 nodes using 32, 64, 128, 256, 1024 and 1536 MPI ranks, respectively. In all our tests, all nodes are fully packed, \emph{i.e.}, one MPI rank is assigned to each core, exerting full load on the memory/cache system. 

\begin{figure}
	\centering
	\includegraphics[width=0.9\columnwidth]{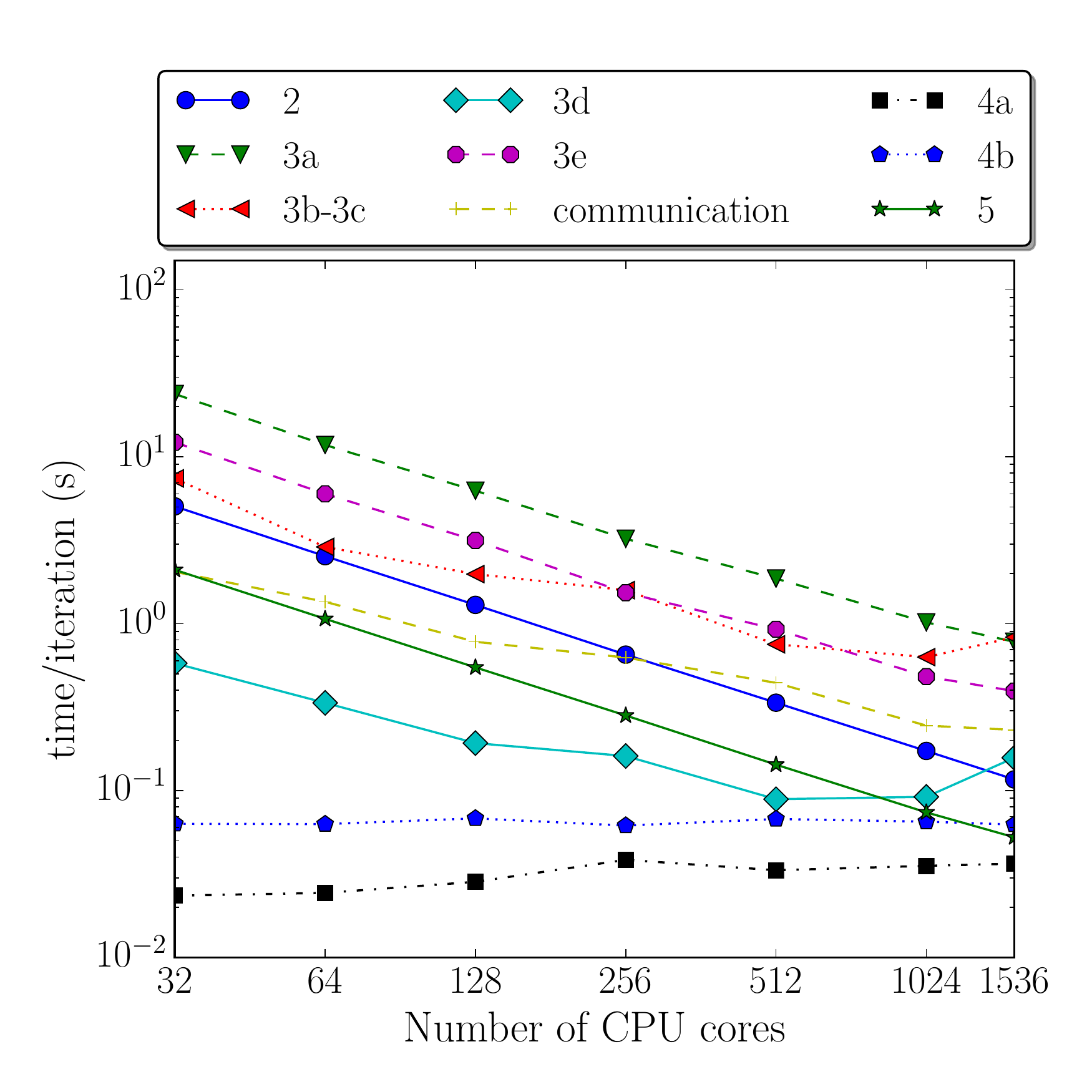}
	\vspace{-0.2in}
	\caption{Scalability of MPI-only version of Sky3D for the 3000 neutron and 3000 proton system using the $L=32\,{\rm fm}$ grid.}
	\vspace{-0.1in}
	\label{fig:3000p_32}
\end{figure}

\begin{figure}
	\centering
	\includegraphics[width=0.9\columnwidth]{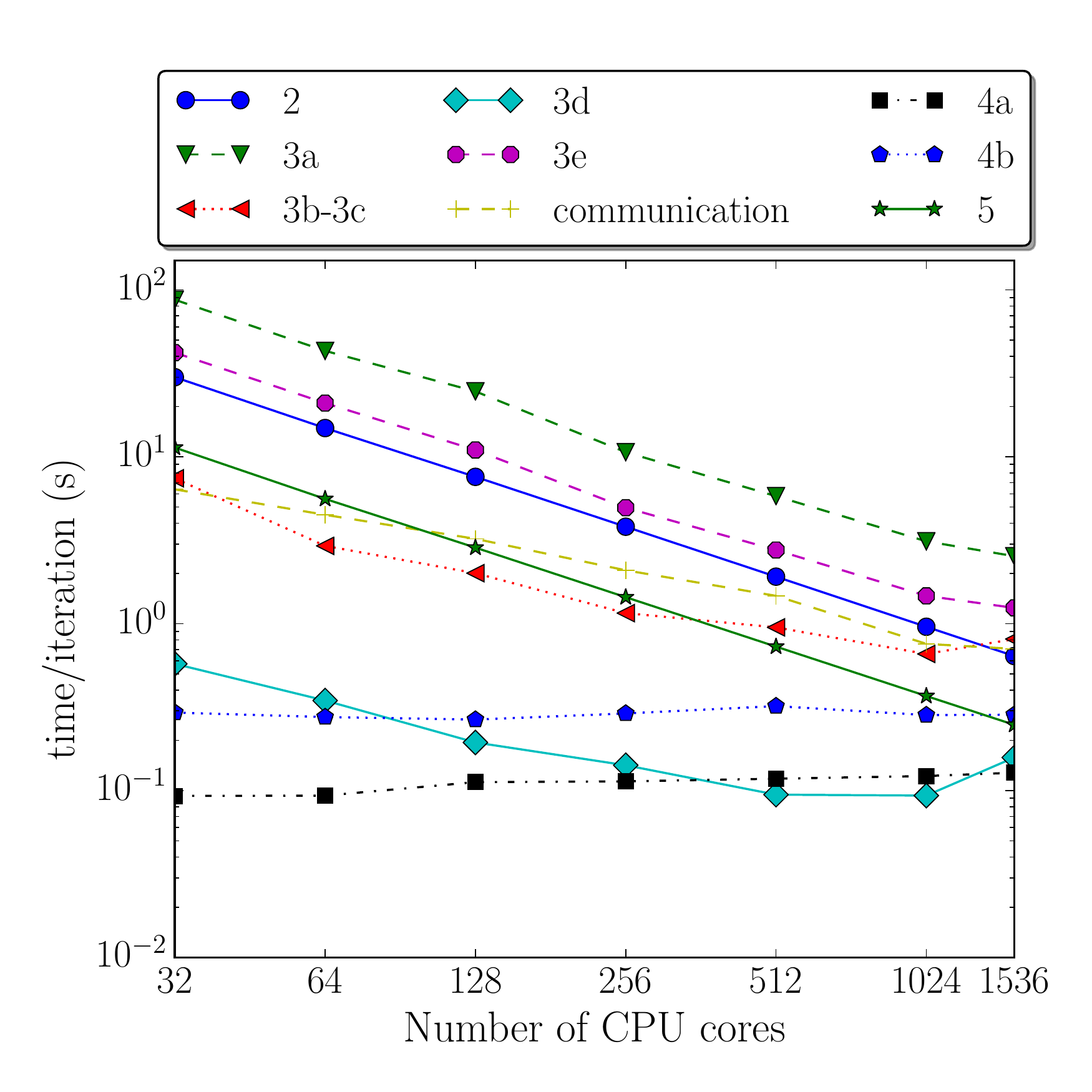}
	\vspace{-0.2in}
	\caption{Scalability of MPI-only version of Sky3D for the 3000 neutron and 3000 proton system using the $L=48\,{\rm fm}$ grid.}
	\vspace{-0.1in}
	\label{fig:3000_48}
\end{figure}

Performance results for $L=32\,{\rm fm}$ and $L=48\,{\rm fm}$ cases are shown in Fig.\,\ref{fig:3000p_32} and Fig.\,\ref{fig:3000_48}, respectively. In both figures, total execution time per iteration is broken down into the time spent for individual steps of Fig.\,\ref{fig:Sky3d_para}. In addition, ``communication" represents the time needed by ScaLAPACK's {\tt PZGEMR2D} routine for data movement during switches between different distributions (\emph{i.e.}, 1D, 2D, and 2D subgroups). For each step, we use a single line to report the time for neutron and proton computations by taking their maximum. 

As seen in Fig.\,\ref{fig:3000p_32}, calculation of the matrices (step 3a) is the most expensive step. Another expensive step is step 3e where diagonalized and orthonormalized wave functions are built. Diagonalization of the Hamiltonian $\mathcal{H}$ and the L\"owdin orthonormalization procedures (steps 3b-3c, which are combined into a single step as they are performed in parallel) also takes significant amount of time. It can be seen that step 4a does not scale well, because it consists mainly of communication of the densities. We also note that step 4b is not parallelized, it is rather performed redundantly on each process because it takes an insignificant amount of time.

The damped gradient step (step 2) and computation of single particle properties (step 5) scale almost perfectly. Steps 3a and 3e, which are compute intensive kernels, also exhibit good scaling. While ScaLAPACK's eigensolver routine {\tt PZHEEVR} performs well for smaller number of cores, it does not scale to a high number of cores. In fact, the internal communications in this routine becomes a significant bottleneck to the extent that steps 3b and 3c become the most expensive part of the calculation on 1536 cores. In Table\,\ref{tab:3000_32}, we give strong scaling efficiencies for the most important parts of the iterations for the $L=32\,{\rm fm}$ grid. Overall, we observe good scaling up to 512 cores, where we achieve 70.5\% efficiency. However, this number drops to 36.1\% on 1536 cores and steps 3b-3c are the main reason for this drop.

\begin{table}
	\centering
	\caption{Scalability of MPI-only version of Sky3D for the $L=32\,{\rm fm}$ grid. Time is given in seconds, and efficiency (eff) is given in percentages.}
	\label{scal_32}
	\renewcommand{\arraystretch}{1.4}
	\tabcolsep=0.1cm
	\begin{tabular}{r |r r|r r|r r|r r}
		\hline
		& \multicolumn{2}{c|}{\textbf{calc. matrix}}    & \multicolumn{2}{c|}{\textbf{recombine}} & \multicolumn{2}{c|}{\textbf{diag+L\"owdin}} & \multicolumn{2}{c}{\textbf{Total}} \\ \hline
		\textbf{cores} & \textbf{time} &    \textbf{eff}  &      \textbf{time}     &   \textbf{eff} &     \textbf{time}      & \textbf{eff}   &    \textbf{time}       &     \textbf{eff}      \\ \hline
		
		32 		& 23.8	& 100 	 & 12.2	& 100   	&     7.4     &    100       &     59.8     &     100      \\ \hline
		64  	& 11.8 	 & 101.5 &  6.0	& 102.0		&    2.9      &     128.7     &   28.4     &     105.1      \\ \hline
		128		& 6.3 	&  94.8  &   3.2  &    96.8       &   2.0     &   93.6       &    16.0      &     93.5     \\ \hline
		256		& 3.2 	& 92.2  &  1.5  &    99.4       &   1.6   &    58.4    &   9.2     &    81.4      \\ \hline
		512		& 1.9 	&  79.8	 &   0.9  &    82.4      &  0.8   &   61.6     &    5.3     &     70.5   \\ \hline
		1024   & 1.0	&  73.0	&  0.5  &    79.1      &    0.6   &    36.8   &   3.4    &     55.0    \\ \hline
		1536   & 0.8	& 63.6 	&  0.4  &      64.4     &   0.8     &   18.6     &   3.5    &    36.1    \\ \hline
	\end{tabular}
	\label{tab:3000_32}
	\vspace{-0.15in}
\end{table}

In Fig.\,\ref{fig:3000_48}, strong scaling results for the $L=48\,{\rm fm}$ grid is shown. In this case, the number of neutrons and protons are the same, but the size of wave functions is larger than the previous case. As a result, the computation intensive steps 3a, 3e and 4b which directly work on these wave functions are significantly more expensive than the corresponding runs for the $L=32\,{\rm fm}$ case. As it is clearly visible in this figure, on small number of nodes, the overall iteration time is dominated by these compute-intensive kernels. This changes in larger scale runs, where the times spent in diagonalization and L\"owdin orthonormalization (steps 3b-3c) along with communication operations also become significant. 

The  increased size of wave functions and increased computational intensity actually results in better scalability. As shown in Table\,\ref{tab:3000_48}, we observe 93.1\% efficiency during matrix construction and 95.5\% efficiency for the recombine step on 1024 cores. We note that the calculate matrix step's efficiency can actually be greater than 100\% owing to the perfect square core counts like 64 and 256 cores. However, like in the previous case, the diagonalization and orthonormalization steps do not scale well for larger number of cores. 
\begin{table}
\centering
\caption{Scalability of MPI-only version of Sky3D for the $L=48\,{\rm fm}$ grid. Time is given in seconds, and efficiency (eff) is given in percentages.}
\label{scal_48}
\renewcommand{\arraystretch}{1.4}
\tabcolsep=0.1cm
\begin{tabular}{r |r r|r r|r r|r r}
\hline
                      & \multicolumn{2}{c|}{\textbf{calc. matrix}}    & \multicolumn{2}{c|}{\textbf{recombine}} & \multicolumn{2}{c|}{\textbf{diag+L\"owdin}} & \multicolumn{2}{c}{\textbf{Total}} \\ \hline
                   \textbf{cores} & \textbf{time} &    \textbf{eff}  &      \textbf{time}     &   \textbf{eff} &     \textbf{time}      & \textbf{eff}   &    \textbf{time}       &     \textbf{eff}      \\ \hline
                   
32 & 87.3 & 100 &     42.1        & 100        &     7.4   &    100       &   192    &     100      \\ \hline
                 64  & 43.1 &       101.2          &     21     &    100.3     &    2.9     &   127     &    95.1    &    100.9    \\ \hline
                128    & 24.7 &   88.5   &  11   &   95.9    &  2.0  & 92.6  &  53.6   & 89.6      \\ \hline
                256      & 10.6 &     102             & 4.9  &   107.2   &  1.2   &   80.3   &    25.51    &   94.1   \\ \hline
                 512     & 5.8 &  94   &  2.8   &   95.2    &  0.9   &  48.9    &   15   &  80.2    \\ \hline
                 1024     & 2.9 &   93.1           &  1.4    &    95.5   &   0.6   &   35.3   &  8.4     &   71.3  \\ \hline
                 1536     & 2.5 &    71.7           &  1.2  &     70.5    &  0.8   &  19.1  &   7.54   &    53    \\ \hline
\end{tabular}
\label{tab:3000_48}
\vspace{-0.15in}
\end{table}

Overall, Sky3D shows good scalability for small to moderate number of nodes, but this decreases slightly with increased core counts. This decrease in efficiency is mainly due to the poor scaling of ScaLAPACK's eigensolver used in the diagonalization and orthonormalization steps, and partially due to the cost of having to communicate large wave functions at each Sky3D iteration.

\subsection{Comparison between MPI-only and MPI/OpenMP Hybrid Parallelization}

\begin{figure}
	\centering
	\includegraphics[width=\columnwidth]{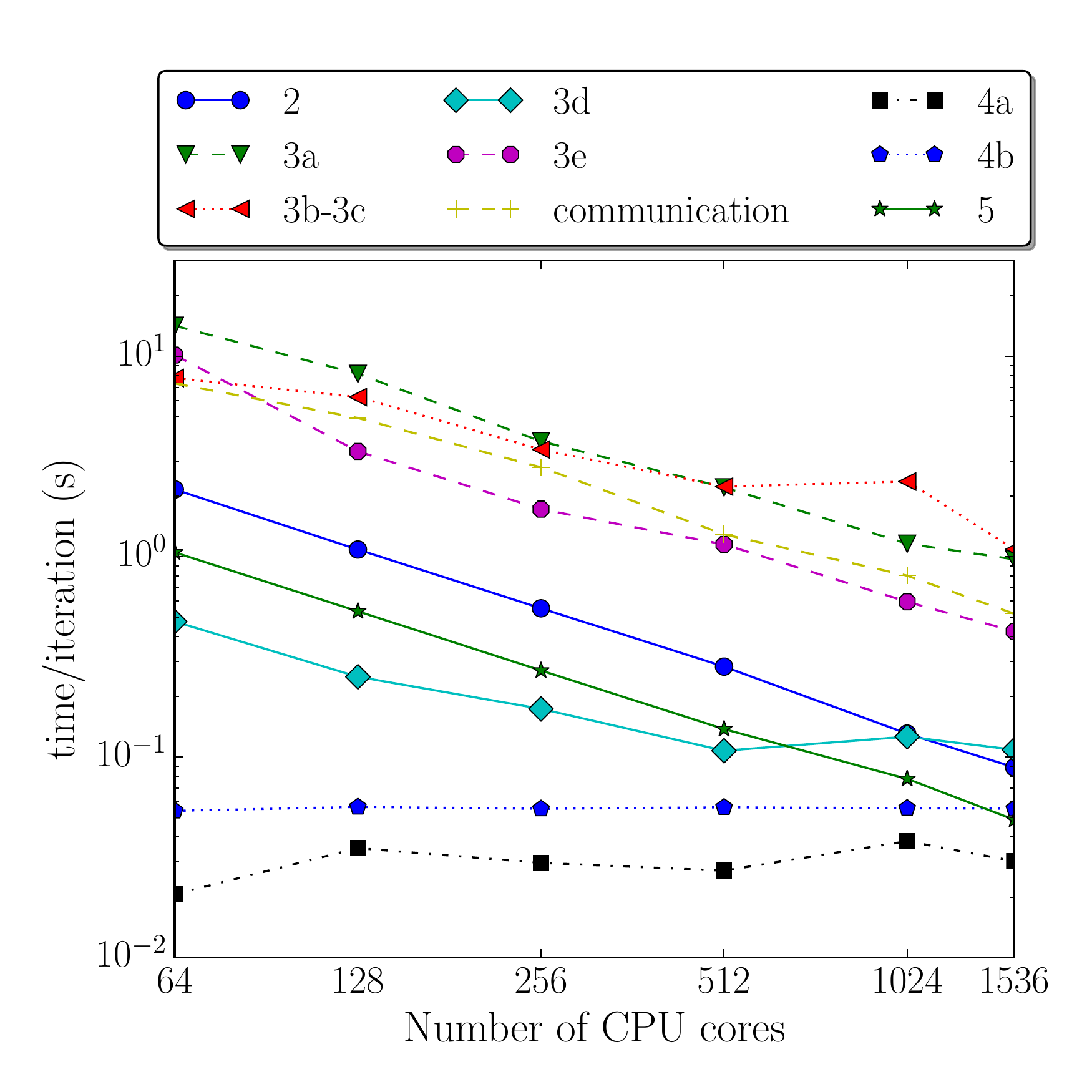}
	\vspace{-0.4in}
	\caption{Scalability of MPI/OpenMP parallel version of Sky3D for the 3000 neutron and 3000 proton system using the $L=32\,{\rm fm}$ grid.}
	\vspace{-0.1in}
	\label{fig:3000_omp_32_numa}
\end{figure}

\begin{figure}
	\centering
	\includegraphics[width=\columnwidth]{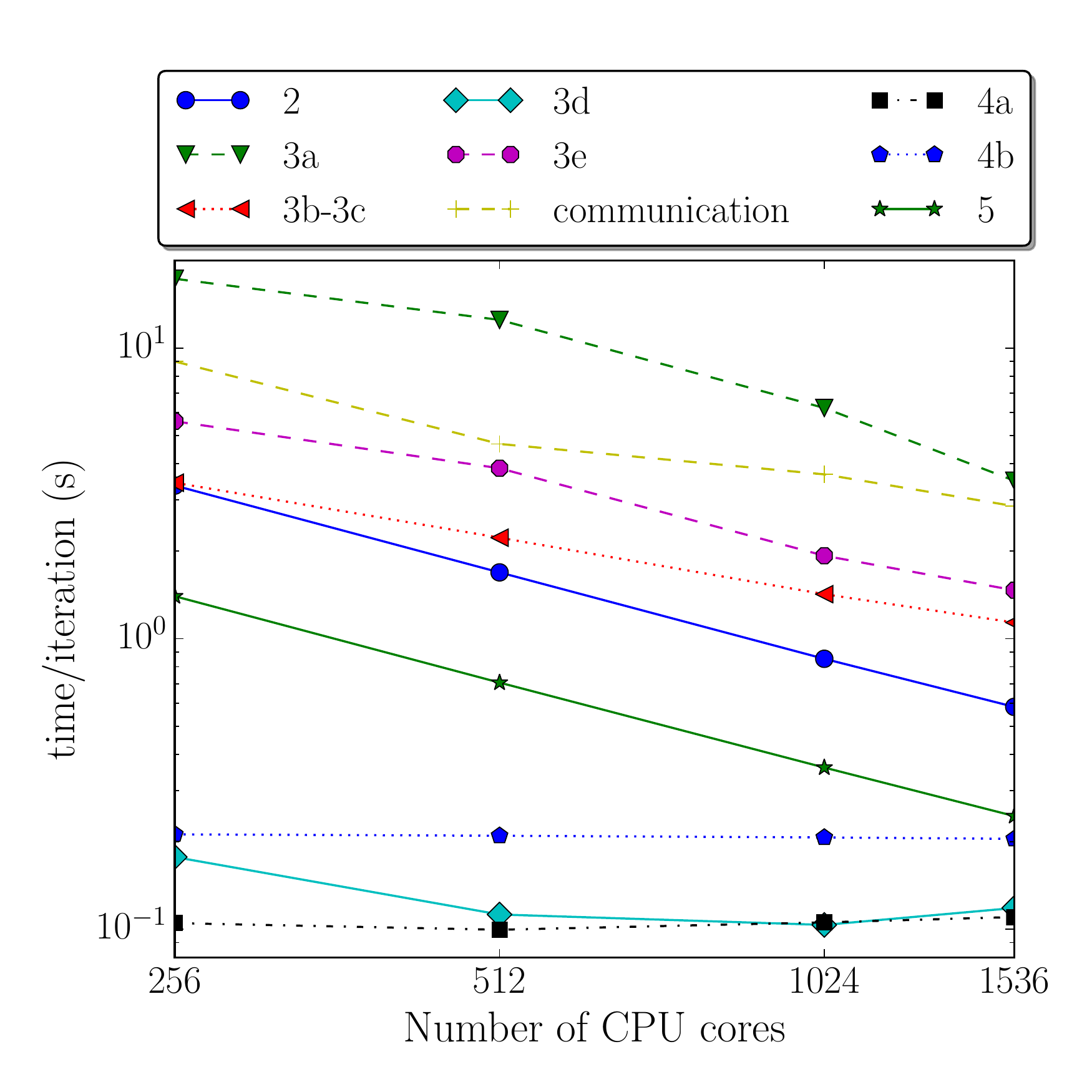}
	\vspace{-0.4in}
	\caption{Scalability of MPI/OpenMP parallel version of Sky3D for the 3000 neutron and 3000 proton system using the $L=48\,{\rm fm}$ grid.}
	\vspace{-0.1in}
	\label{fig:3000_omp_48}
\end{figure}

On Cori, "Haswell" compute nodes contain two sockets with 16 cores each. To prevent any performance degradations due to non-uniform memory accesses (NUMA), we performed our tests using 2 MPI ranks per node with each MPI rank having 16 OpenMP threads executed on a single socket. 
Since we are grouping the available cores into neutron and proton groups which are further divided in half for running diagonalization and orthonormalization tasks in parallel, we need a minimum of 4 MPI ranks in each test. In Figures \ref{fig:3000_omp_32_numa} and \ref{fig:3000_omp_48}, we show strong scalability test results similar to the MPI-only implementation discussed earlier. In this case, the legends along the x-axis denotes the total core counts. For example, 128 means that we are running this test on 4 nodes with 8 MPI ranks and 16 OpenMP threads per rank. For the $L=32\,{\rm fm}$ grid, we have tested the MPI/OpenMP hybrid parallel version with 4, 8, 16, 32, 64, 96 MPI ranks. For the $L=48\,{\rm fm}$ grid, we have a larger number of wave functions for which  ScaLAPACK's data redistribution routine {\tt PZGEMR2D} runs out of memory on low node counts. Hence, we tested this case with 16, 32, 64 and 96 MPI ranks only.

In Fig.\,\ref{fig:3000_omp_32_numa}, we show the strong scaling results for the $L=32\,{\rm fm}$ case, along with detailed efficiency numbers for the computationally expensive steps in Table\,\ref{tab:3000_32_omp}. Similar to the MPI-only case, the compute-intensive matrix construction and recombine phases show good scalability, but the diagonalization and L\"owdin orthonormalization part does not perform as well as these two parts. While the strong scaling efficiency numbers in this case look better than the MPI-only case (see Table\,\ref{tab:3000_32}), we note that this is due to the inferior performance of the MPI/OpenMP parallel version for its base case of 64 cores. In fact, the recombine part performs so poor on 64 cores that its strong scaling efficiency is constantly over 100\% almost all the way up to 1536 cores. But comparing the total execution times, we see that the MPI-only code takes 28.4 seconds on average per iteration, while the MPI/OpenMP parallel version takes 50.7 seconds for this same problem on 64 cores. 

\begin{table}
	\centering
	\caption{Scalability of MPI/OpenMP parallel version of Sky3D for the $L=32\,{\rm fm}$ grid. Time is given in seconds, and efficiency (eff) is given in percentages.}
	\label{scal_32_omp}
	\renewcommand{\arraystretch}{1.4}
	\tabcolsep=0.11cm
	\begin{tabular}{r |r r|r r|r r|r r}
		\hline
		& \multicolumn{2}{c|}{\textbf{calc. matrix}}    & \multicolumn{2}{c|}{\textbf{recombine}} & \multicolumn{2}{c|}{\textbf{diag+lowedin}} & \multicolumn{2}{c}{\textbf{Total}} \\ \hline
		\textbf{cores} & \textbf{time} &    \textbf{eff}  &      \textbf{time}     &   \textbf{eff} &     \textbf{time}      & \textbf{eff}   &    \textbf{time}       &     \textbf{eff}      \\ \hline
		
		64    & 14.2 &   100    &    10.1   &    100    &  7.8  &      100    &  50.7   &    100   \\ \hline
		128    & 8.2 &  86.8    &    3.34    &  151.2       & 6.2    &    62.3   &    30.7   &   82.6   \\ \hline
		256      & 3.8 &     94.2       &  1.7  & 146.8      & 3.4 & 56.8  &  16   &   79.4   \\ \hline
		512     & 2.2 &  80  & 1.1 &  110.1      & 2.2  &  43.5   &  9.6   & 66  \\ \hline
		1024     & 1.2 &       76.4           & 0.6 &   106.3      &   2.4   & 20.5  &   6.8   &   46.7    \\ \hline
		1536     & 1 &     61        & 0.4 &   99.7   &  1.1  &   29.8  & 4.4  &  48.2   \\ \hline		
	\end{tabular}
	\label{tab:3000_32_omp}
	\vspace{-0.15in}
\end{table}

\begin{figure}
	\centering
	\includegraphics[width=\columnwidth]{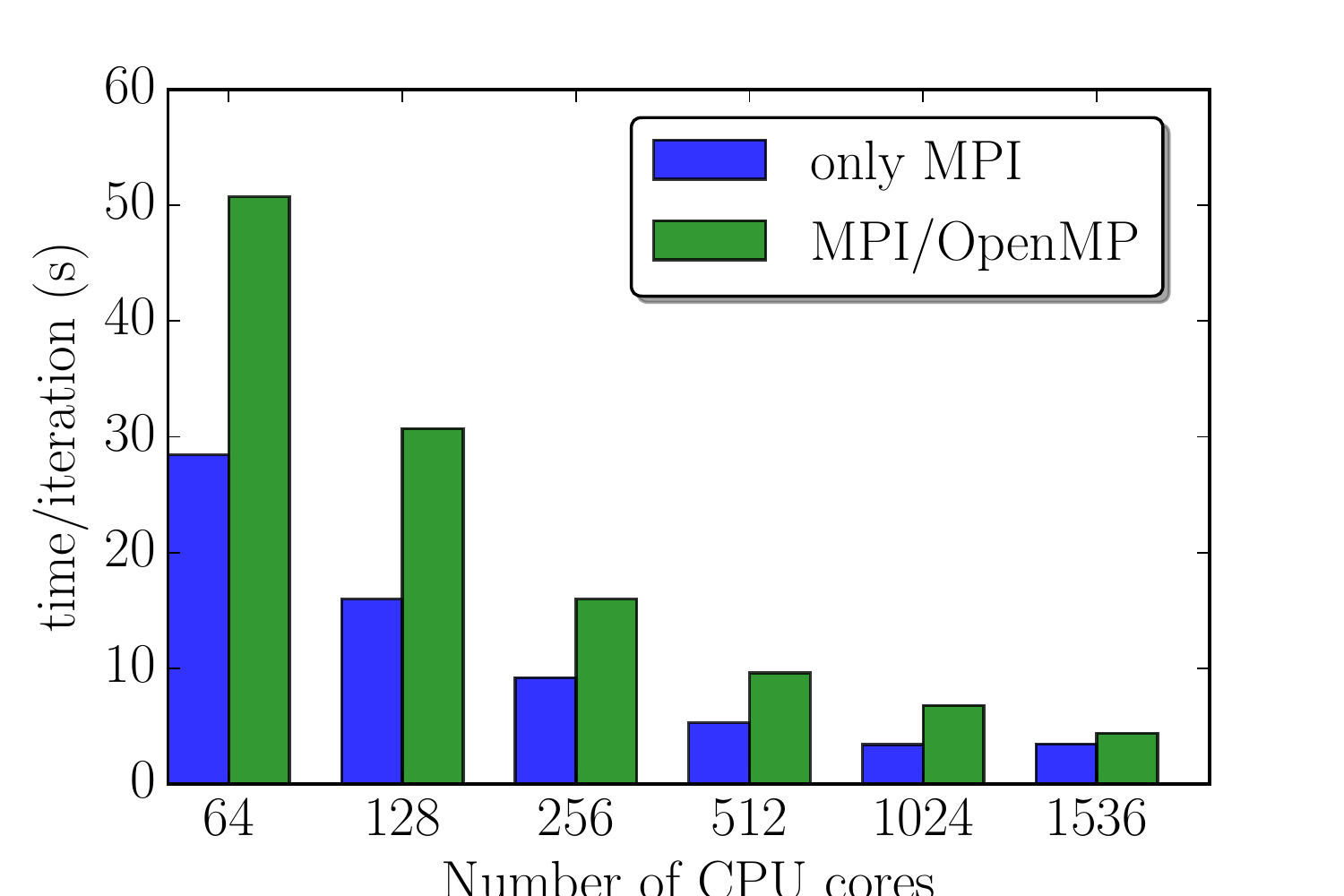}
	\vspace{-0.3in}
	\caption{Comparison of the execution times for the MPI-only and MPI/OpenMP parallel versions of Sky3D for the 3000 neutron and 3000 proton system using the $L=32\,{\rm fm}$ grid. }
	\label{fig:combine_32}
	\vspace{-0.1in}
\end{figure}

This performance issue in the MPI/OpenMP version persists through all test cases as shown in Figure\,\ref{fig:combine_32}. In particular, for smaller number of cores, the performance of MPI/OpenMP version is poor compared to the MPI-only version. With increasing core counts, the performance difference lessens slightly, but the MPI-only version still performs better. In general, we observe that the MPI-only version outperforms the MPI/OpenMP version by a factor of about 1.5x to 2x. The main reason behind this is that the thread parallel ScaLAPACK routines used in the MPI/OpenMP implementation perform worse than the MPI-only ScaLAPACK routines, which is contrary to what one might naively expect, given that our tests are performed on a multi-core architecture.

\begin{figure}
\centering
\includegraphics[width=\columnwidth]{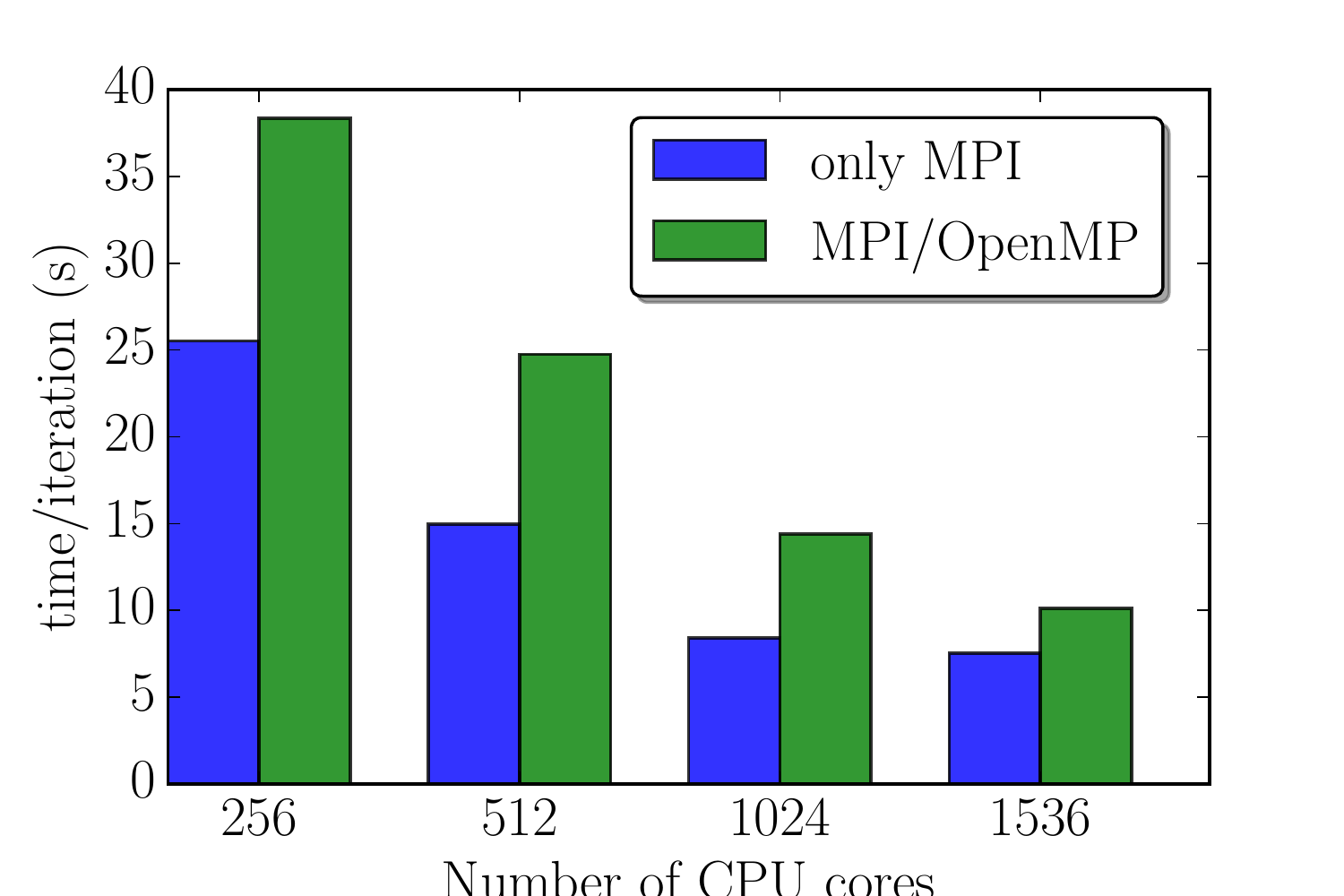}
\vspace{-0.20in}
\caption{Comparison of the execution times for the MPI-only and MPI/OpenMP parallel versions of Sky3D for the 3000 neutron and 3000 proton system using the $L=32\,{\rm fm}$ grid. }
\vspace{-0.1in}
\label{fig:combine_48}
\end{figure}

In Fig.\,\ref{fig:3000_omp_48} and Table\,\ref{tab:3000_48_omp}, we show the strong scaling results for the $L=48\,{\rm fm}$ grid. Again, in this case parts directly working with the wave functions, \emph{i.e.}, calculation of matrices (step 3a) and building of orthonormalized and diagonalized wave functions (step 3e), become significantly more expensive compared to the diagonalization and L\"owdin orthonormalizations (steps 3b \& 3c). Of particular note here is the more pronounced communication times during switches between different data distributions which is mainly due to the larger size of the wave functions. Overall, we obtain 64.5\% strong scaling efficiency using up to 96 MPI ranks with 16 threads per rank (1536 cores in total). In terms of total execution times though, MPI/OpenMP parallel version still underperforms compared to the MPI-only version (see Fig.\,\ref{fig:combine_48}).

\begin{table}
\centering
\caption{Scalability of MPI/OpenMP parallel version of Sky3D for the $L=48\,{\rm fm}$ grid. Time is given in seconds, and efficiency (eff) is given in percentages.}
\label{scal_48_omp}
\renewcommand{\arraystretch}{1.4}
\tabcolsep=0.1cm
\begin{tabular}{r |r r|r r|r r|r r}
\hline
                      & \multicolumn{2}{c|}{\textbf{calc. matrix}}    & \multicolumn{2}{c|}{\textbf{recombine}} & \multicolumn{2}{c|}{\textbf{diag+L\"owdin}} & \multicolumn{2}{c}{\textbf{Total}} \\ \hline
                   \textbf{cores} & \textbf{time} &    \textbf{eff}  &      \textbf{time}     &   \textbf{eff} &     \textbf{time}      & \textbf{eff}   &    \textbf{time}       &     \textbf{eff}      \\ \hline
                   
                256      & 17.3 &   100    & 5.6  &    100    &  3.4 &  100   & 43.7  &   100    \\ \hline
                 512     & 12.5 &  69.3  & 3.9 &  72.5   & 2.2 & 77.3   &  28.1   &  77.9 \\ \hline
                 1024     & 6.2 &   69.6   & 1.9 &  72.5   &  1.4  & 60.5 &  16.2  &   67.4     \\ \hline
                 1536     & 3.5 &   82.5      &  1.5  & 63.6   &  1.1   &   50.5    & 11.3   &  64.5   \\ \hline
\end{tabular}
\label{tab:3000_48_omp}
\vspace{-0.15in}
\end{table}

\subsection{Load Balancing}

\begin{figure}
	\centering
	\includegraphics[width=0.9\columnwidth]{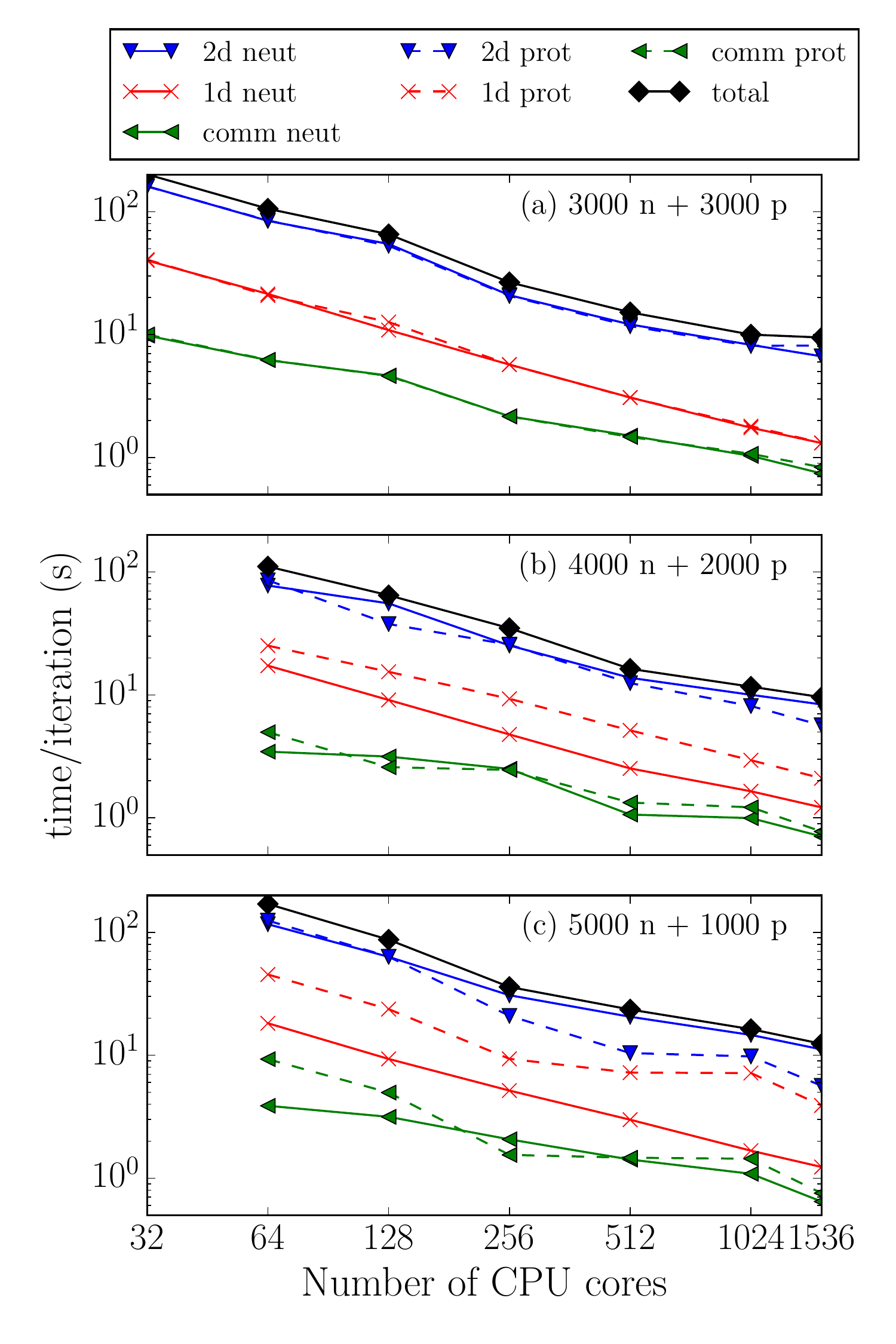}
	\vspace{-0.2in}
	\caption{Times per iteration for neutron and proton processor groups, illustrating the load balance for the 3000 neutrons and 3000 protons (a), 4000 neutrons and 2000 protons (b), and 5000 neutrons and 1000 protons (c) systems using the $L=48\,{\rm fm}$ grid. Due to memory constraints the latter two cases cannot be calculated using 32 CPU.}
	\vspace{-0.1in}
	\label{fig:3-4-5_lb_short}
\end{figure}

In this section, we analyze the performance of our load balancing approach which divides the available cores into neutron and proton groups for parallel execution. For better presentation, we break down the execution time into three major components: Calculations in steps using a 2D data distribution, calculations using a 1D distribution of wave functions and communication times. In Fig.~\ref{fig:3-4-5_lb_short}(a), we show the time taken by the cores in the neutron and proton groups for the 3000 neutron and 3000 proton system using the $L=48\,{\rm fm}$ grid - which is essentially the same plot as in the previous section, but it gives the timings for neutrons and protons separately. As this system has an equal number of neutrons and protons, available cores are divided equally into two groups. As can be seen in Fig.~\ref{fig:3-4-5_lb_short}(a), the time needed for different steps in this case is almost exactly identical for neutrons and protons.


In Figure\,\ref{fig:3-4-5_lb_short}(b), we present the results for a system with 4000 neutrons and 2000 protons. In this case, according to our load balancing scheme, the number of cores in the neutron group will be roughly 4x larger than the number of cores in the proton group because we distribute the cores based on the ratio of the square of the number of particles in each group. We observe that all three major parts are almost equally balanced for up to 1024 cores, but 2D calculations for neutrons is slightly more expensive on 1536 cores. A more detailed examination reveals that this difference is due to the eigendecomposition times in steps 3b-3c. However, it is relatively minor compared to the total execution time per iteration.

Note that the times for the steps with 1d distribution show some variation for the system with 4000 neutrons and 2000 protons. This is due to the fact that we split the available cores into neutron and proton groups based on the cost of steps with 2D data distributions. Consequently, 1D distributed steps take more time on the proton processor group, but this difference is negligible in comparison to the cost of 2D distributed steps.

In Figure~\ref{fig:3-4-5_lb_short}(c) results for a more challenging case with 5000 neutrons and 1000 protons are presented. In this case the majority of the available cores are assigned to the neutron group - more precisely, the ratio between the sizes of the two groups is roughly 25. We observe that 1D calculations take significantly more time for protons in this case, but any potential load imbalances are compensated by the reduced 2D calculation times for protons. 

\begin{figure}
	\centering
	\includegraphics[width=0.9\columnwidth]{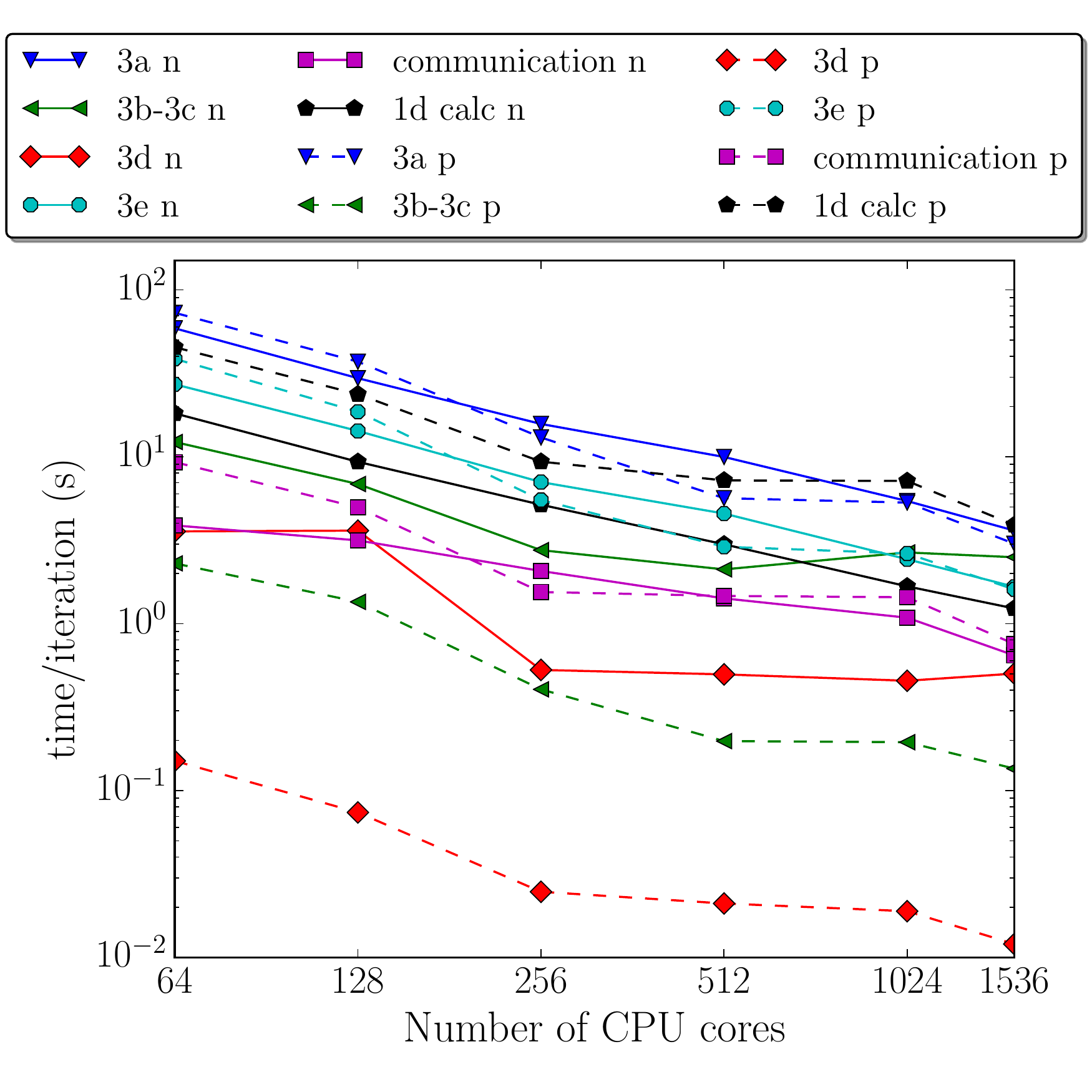}
	\vspace{-0.2in}
	\caption{A detailed breakdown of per iteration times for neutron and proton processor groups, illustrating the load balance for the 5000 neutrons and 1000 protons system.}
	\vspace{-0.1in}
	\label{fig:5000_lb}
\end{figure}
A further inspection of the execution time of each step for the 5000 neutron and 1000 proton system is given in Figure~\ref{fig:5000_lb}. This inspection reveals that time needed for neutrons and protons mainly differ for step 3b-3c and step 3d due to the large difference between neutron and proton counts. But these difference are not significant compared to the other computationally heavy steps which are well load balanced. As shown in Table\,\ref{tab:5000_48}, our implementation still achieves about 50\% strong scaling efficiency on 1536 cores for this challanging case with 5000 neutrons and 1000 protons. 

\begin{table}
	\centering
	\caption{Scalability of MPI-only version of Sky3D for the 5000 neutrons and 1000 protons system using  the $L=48\,{\rm fm}$ grid. Time is given in seconds, and efficiency (eff) is given in percentages.}
	\label{scal_32_5000}
	\renewcommand{\arraystretch}{1.4}
	\tabcolsep=0.09cm
	\begin{tabular}{r |r r|r r|r r|r r}
		\hline
		& \multicolumn{2}{c|}{\textbf{calc. matrix}}    & \multicolumn{2}{c|}{\textbf{recombine}} & \multicolumn{2}{c|}{\textbf{diag+L\"owdin}} & \multicolumn{2}{c}{\textbf{Total}} \\ \hline
		\textbf{cores} & \textbf{time} &    \textbf{eff}  &      \textbf{time}     &   \textbf{eff} &     \textbf{time}      & \textbf{eff}   &    \textbf{time}       &     \textbf{eff}      \\ \hline
		
		64 & 72.9 & 100 & 38.6  & 100        &    2.3   &    100       &    170    &     100      \\ \hline
		128    & 37.1 &  98.1   &   18.6      &  103.4     &   1.35  & 85      &  87   &  97.6    \\ \hline
		256      & 15.7 &       115.8             & 7.1  &    136.9     & 2.75 &  20.82      &    35.9    &   118.3   \\ \hline
		512     & 10  & 91.3 &    4.6    &   105.7    &  2.1   &13.6  &      23.5  &   90.3  \\ \hline
		1024     & 5.4 &     83.9            &  2.4   &   99.2     &   2.7   &  5.3   &   16.3    &    65.2  \\ \hline
		1536     & 3.6 &      84   &  1.7   &  96.7  &  2.5  &  3.8   &     12.4 &   57.2   \\ \hline
	\end{tabular}
	\label{tab:5000_48}
	\vspace{-0.1in}
\end{table}

\section{Conclusions and Future Work}
In this paper, we described efficient and scalable techniques used to parallelize Sky3D, a nuclear DFT solver that operates on an equidistant grid in a pure MPI framework as well as a hybrid MPI/OpenMP framework. By carefully analyzing the computational motifs in each step and data dependencies between different steps, we used a 2D decomposition scheme for Sky3D kernels working with matrices, while using a 1D scheme for those performing computations on wave functions. We presented load balancing techniques which can efficiently leverage high degrees of parallelism by splitting available processors into neutron and proton groups. We also presented algorithmic techniques that reduce the total execution time by overlapping diagonalization and orthogonalization steps using  subgroups within each processor group. Detailed performance analysis on a multi-core architecture (Cori at NERSC) reveal that parallel Sky3D can achieve good scaling to a moderately large number of cores. Contrary to what one might naively expect, the MPI-only implementation outperforms the hybrid MPI/OpenMP implementation, mainly because ScaLAPACK's eigedecomposition routines perform worse in the hybrid parallel case. For larger core counts, the disparity between the two implementations seems to be less pronounced. As a result of detailed performance evaluations, we have observed that 256 to 1024 processors are reasonably efficient for nuclear pasta simulations and we consider these core counts for production runs, depending on the exact calculation.

Using the new MPI parallel Sky3D code, we expect that pasta phases can be calculated for over 10,000 nucleons in a fairly large box using a quantum mechanical treatment. As a result, we expect to reach an important milestone in this field. We plan to calculate properties of more complicated pasta shapes and investigate defects in pasta structures which occur in large systems.

As part of future work, we plan to extend our implementation and optimize it for Xeon Phi and GPU architectures, too. While acceleration on Xeon Phi or GPUs may generally be a major task for scientific codes, we anticipate that this work will be relatively less challenging, as a result of our decision to base Sky3D on well-optimized ScaLAPACK and FFTW libraries.

\section*{Acknowledgment}
This work was supported by a startup grant from the College of Engineering at Michigan State University and by the U.S. Department of Energy, under Award Numbers DOE-DE-NA0002847 (NNSA, the Stewardship Science Academic Alliances program) and DE-SC0013365 (Office of Science). The authors would like to thank the developers of the Sky3D. In particular B.S. would like to thank P.-G. Reinhard, J.A. Maruhn and S. Umar for the very useful discussions and W. Nazarewicz for the support provided.

\section*{References}
\bibliographystyle{IEEEtran}

\bibliography{Physics,cs}

\end{document}